\documentclass[
  aps,
  reprint,
  prd,
  amsmath,
  amssymb,
  nofootinbib,
  showkeys,
  longbibliography
]{revtex4-2}

\usepackage{graphicx}
\usepackage{bm}
\usepackage{amsfonts}
\usepackage{slashed}
\usepackage{dcolumn}

\usepackage{xcolor}
\usepackage{soul}
\soulregister\cite7
\soulregister\ref7
\soulregister\eqref7
\soulregister\textit7
\soulregister\emph7

\begin{document}
\title{Semiclassical phases of charged spin-$\tfrac{1}{2}$ matter-wave interferometers\\
in gravitational-wave backgrounds}

\author{Nontapat Wanwieng}
\email{nontapat@narit.or.th}
\affiliation{National Astronomical Research Institute of Thailand (Public Organization), Chiang Mai 50180, Thailand}

\author{Apimook Watcharangkool}
\affiliation{National Astronomical Research Institute of Thailand (Public Organization), Chiang Mai 50180, Thailand}

\date{\today}

\begin{abstract}
A matter wave propagating through curved spacetime accumulates a phase
that encodes both geometry and gauge structure. We develop a
semiclassical description of charged spin-$\tfrac{1}{2}$ matter-wave
interferometry based on a WKB expansion of the general covariant Dirac equation.
At leading order, and within the guided-arm approximation, the
interferometric phase separates into three additive contributions: a
dynamical phase associated with proper-time evolution, a spin phase
arising from parallel transport in the local Lorentz frame, and an
electromagnetic phase generated by curvature-induced perturbations of a
background electromagnetic field. In a freely falling detector frame
described by Fermi normal coordinates, the dynamical and spin channels
are driven by the local gravitoelectric and gravitomagnetic tidal
fields, while the electromagnetic channel follows from solving the
detector-frame Maxwell equations for a uniform background magnetic
field in the long-wavelength regime. For a weak gravitational wave, all
three channels are controlled by the common tidal scale
$\ddot h_{+,\times}\sim\Omega_{\rm gw}^2 h_0$, but enter through
distinct couplings and geometry-dependent response kernels. Applied to
an idealized square Mach--Zehnder interferometer, the dynamical channel reproduces
the established tidal matter-wave response, and the spin channel is
suppressed by the ratio of the Compton wavelength to the interferometer
size. The electromagnetic channel is set by the applied magnetic flux
rather than by the particle mass or spin: within the initial-value
solution adopted here, its magnitude depends on the orientation of the
background field relative to the wave, with the perpendicular
configuration carrying an additional factor $c/v$, and the two
orientations select different gravitational-wave polarizations.
\end{abstract}

\maketitle
\section{Introduction}
\label{sec:intro}

Quantum interference provides a direct probe of gauge fields and spacetime
geometry through the phase accumulated by a matter wave propagating in
external fields or curved spacetime. A paradigmatic example is the
Aharonov--Bohm (AB) effect~\cite{Aharonov1959}, in which a charged particle
acquires a phase from the electromagnetic vector potential even when
traversing regions where the field strength vanishes, {establishing the
physical significance of gauge potentials in quantum mechanics.} More
generally, geometric phases arise from cyclic evolution in parameter
space~\cite{Berry1984,AharonovAnandan1987} and have been confirmed in
electron interferometry~\cite{Tonomura1986}.

A complementary manifestation arises in gravitation. The
Colella--Overhauser--Werner (COW) experiment~\cite{colella1975} demonstrated
that a gravitational field induces a phase shift in a matter-wave
interferometer, understood relativistically as a proper-time difference
accumulated along distinct trajectories~\cite{Stodolsky1979}. Beyond uniform
acceleration, large-separation atom interferometers resolve tidal phase
shifts produced by spacetime curvature across the wavefunction
~\cite{Asenbaum2017}, {and differential acceleration measurements at the
$10^{-12}$ level have been achieved in equivalence-principle
tests~\cite{Asenbaum2020}.} Atom interferometry has correspondingly emerged
as a platform for precision tests of gravity and gravitational-wave (GW)
detection~\cite{Dimopoulos2008,Graham2013}, with long-baseline concepts
based on laser-phase measurements, atomic clocks, and inertial sensing under
development~\cite{Canuel2018,MAGIS2021}.

{On the theoretical side,} early work by Anandan developed a
relativistic description of quantum interference in non-inertial frames
consistent with the equivalence principle~\cite{Anandan1977,Anandan1981}.
Subsequent studies investigated dynamical phases~\cite{Stodolsky1979}, spin-gravity couplings and geometric spin phases~\cite{Mashhoon1988,Papini2008}, {including a unified
relativistic treatment of spin-gravitation phase shifts for Dirac
particles in matter-wave interferometers~\cite{Borde2000} and, more
recently, a framing of GW-induced spin rotation as an internal
Lorentz-group holonomy in Sagnac interferometry~\cite{Javadinezhad2026},}
relativistic spin transport in curved
spacetime~\cite{Audretsch1981}, and electromagnetic responses to
gravitational perturbations within detector-frame formulations of Maxwell's
equations~\cite{Marklund2000,Brodin2001,Servin2003,Domcke2022,Ruggiero2025}. {More recently, relativistic analyses have clarified the response of
atom interferometers and atomic sensors to linearized gravitational
perturbations, including Doppler, Shapiro, and Einstein propagation phases,
laboratory-frame descriptions, and tidal-Hamiltonian formulations in Fermi
normal coordinates~\cite{Badurina2025,Schaffrath2025,Gaudio2026}. These
mechanisms, however, are generally treated within separate theoretical
frameworks: a derivation of the dynamical, spin, and electromagnetic phase
contributions from a single relativistic wave equation in a time-dependent
spacetime has not, to our knowledge, been given. The question is
particularly relevant for GWs, whose time-dependent tidal fields affect
both the proper-time evolution of matter waves and the electromagnetic
environment through which they propagate.}

{
In this work we develop such a description for charged
spin-$\tfrac{1}{2}$ matter waves, based on a WKB expansion of the covariant
Dirac equation. In the detector frame the interferometric phase separates
into three contributions: a dynamical phase associated with proper-time
evolution, a spin-transport phase generated by the local Lorentz
connection, and an electromagnetic phase arising from curvature-induced
perturbations of a background electromagnetic field. Each mechanism has
been studied previously in complementary contexts
~\cite{Stodolsky1979,Mashhoon1988,Borde2000,Javadinezhad2026,Badurina2025,Schaffrath2025,Gaudio2026};
here all three are obtained within a common treatment, allowing their
physical origin and relative scaling to be compared directly.
}

The theory is formulated in a general detector frame with arbitrary
acceleration and rotation, where spacetime curvature is encoded in local
tidal tensors; for GW applications we specialize to a freely falling frame
described by Fermi normal coordinates
(FNC)~\cite{ManasseMisner1963,MTW,PoissonPoundVega2011}. {Within this
frame we reduce the curved-spacetime Maxwell equations in the
long-wavelength limit and solve for the perturbation of a uniform
background magnetic field induced by a weak GW. In the presence of such a
background flux, the induced perturbation generates a time-dependent,
gauge-invariant electromagnetic phase, which admits a four-dimensional
Stokes representation. Because the interferometer is immersed in a
nonvanishing field, this phase is distinct from the canonical field-free
AB effect and is interpreted as a curvature-induced electromagnetic
phase.}

{
Applying the formalism to a square Mach--Zehnder interferometer, we derive
the response of each channel to a monochromatic GW. The dynamical channel
reproduces the standard tidal matter-wave response; the spin channel probes
the gravitomagnetic sector but is suppressed by the ratio of the Compton
wavelength to the interferometer size, and is included primarily for
conceptual completeness. The electromagnetic channel exhibits two features
of note: its amplitude is set by the applied magnetic flux rather than by
the particle mass, and its magnitude and polarization sensitivity depend
on the orientation of the background field relative to the GW propagation
direction. Representative order-of-magnitude estimates illustrate the
resulting hierarchy of the three channels.
}

The paper is organized as follows. Section~\ref{sec:quantumphase} develops
the semiclassical expansion of the Dirac equation and the phase
decomposition. Section~\ref{sec:decomp} reformulates the results in the
detector frame. Section~\ref{sec:AB_GW} {reduces the detector-frame
Maxwell equations and solves for the GW-induced electromagnetic response.}
Section~\ref{sec:interferometric_response} presents the interferometric
response of a square Mach--Zehnder interferometer together with
representative estimates. Section~\ref{sec:conclusion} summarizes our
conclusions.

\section{Semiclassical framework for matter-wave phases}
\label{sec:quantumphase}

A massive spin-$\tfrac12$ Dirac field $\psi(x)$ of mass $m$ and charge $q$
propagates on a curved spacetime with metric signature
$(-,+,+,+)$ and is minimally coupled to an external electromagnetic
potential $A_\mu$. Its dynamics are governed by the covariant Dirac
equation~\cite{collas2019,ParkerToms2009}
\begin{equation}\label{eq:DiracCurved}
{
\left[
\gamma^\mu(x)\mathcal D_\mu
+
\frac{mc}{\hbar}
\right]\psi(x)=0,
}
\end{equation}
with covariant derivative
\begin{equation}
\mathcal D_\mu
=
\partial_\mu+\Gamma_\mu
-\frac{iq}{\hbar}A_\mu .
\end{equation}
The spacetime-dependent gamma matrices are defined by
$\gamma^\mu(x)=e^\mu{}_a(x)\gamma^a$, where $\gamma^a$ are constant Dirac
matrices in the local Lorentz frame and satisfy
\begin{equation}
\{\gamma^\mu(x),\gamma^\nu(x)\}
=
2g^{\mu\nu}(x)\mathbb I .
\end{equation}
{In the local frame the constant matrices obey
$\{\gamma^a,\gamma^b\}=2\eta^{ab}\mathbb I$ with
$\eta^{ab}=\mathrm{diag}(-1,+1,+1,+1)$, together with the hermiticity
assignment $(\gamma^0)^\dagger=-\gamma^0$ and
$(\gamma^k)^\dagger=+\gamma^k$; Eq.~\eqref{eq:DiracCurved} is then the
mostly-plus form of the Dirac equation, equivalent to the familiar
$(i\gamma^\mu\mathcal D_\mu-mc/\hbar)\psi=0$ of the mostly-minus
convention.}
The spin connection is
\begin{equation}\label{eq:Gamma_mu}
{
\Gamma_\mu
=
-\frac{i}{4}\omega_{\mu ab}\Sigma^{ab},
}
\qquad
\Sigma^{ab}
=
\frac{i}{2}[\gamma^a,\gamma^b].
\end{equation}

The phases accumulated by a matter wave follow from
Eq.~\eqref{eq:DiracCurved} in the semiclassical regime, where the
background geometry and external fields vary slowly compared with the
local de Broglie wavelength. In this regime, the phase structure emerges
from a WKB expansion, following the semiclassical analysis of Dirac
fields in curved spacetime~\cite{Oancea2023}. The field is written as
\begin{equation}\label{eq:eikonal}
\psi(x) =
\left(a_0(x) + \hbar a_1(x) + \mathcal{O}(\hbar^2)\right)
\exp\!\left(\frac{i}{\hbar}S(x)\right),
\end{equation}
where $S(x)$ is real at leading order; subleading imaginary contributions
are absorbed into the amplitude.

Substitution of Eq.~\eqref{eq:eikonal} into Eq.~\eqref{eq:DiracCurved}
and expansion in powers of $\hbar$ yield, at leading order,
\begin{equation}
{\left(i\gamma^\mu \Pi_\mu + mc\right)a_0 = 0,}
\end{equation}
with the kinetic momentum $\Pi_\mu \equiv \partial_\mu S - qA_\mu$.
{Under a $U(1)$ gauge transformation,
$A_\mu\rightarrow A_\mu+\partial_\mu\Lambda$,
the eikonal phase transforms as $S\rightarrow S+q\Lambda$,
so that $\Pi_\mu$ remains unchanged.}

{Acting with the conjugate operator
$(-i\gamma^\nu\Pi_\nu+mc)$ and using the Clifford algebra,
nontrivial solutions require}
\begin{equation}\label{eq:HJ}
g^{\mu\nu}\Pi_\mu\Pi_\nu + m^2c^2 = 0,
\end{equation}
which is the relativistic Hamilton--Jacobi equation.
{Its integral curves satisfy $\Pi^\mu=mu^\mu$, with
$u^\mu=dx^\mu/d\tau$ and $u_\mu u^\mu=-c^2$.} In the absence of
electromagnetic fields they reduce to timelike geodesics; in general,
they describe motion under the Lorentz force. At leading order the wave
packet follows these classical trajectories.

{The semiclassical phase accumulated along a world line $\gamma$
connecting the source event to the detection event is}
{
\begin{equation}
\phi_\gamma
=
\frac{1}{\hbar}\int_\gamma \partial_\mu S\, dx^\mu
=
\frac{1}{\hbar}\int_\gamma \left(\Pi_\mu + qA_\mu\right) dx^\mu ,
\label{eq:S_split}
\end{equation}
which, using $\Pi_\mu dx^\mu = m u_\mu u^\mu\,d\tau = -mc^2\,d\tau$,
decomposes into dynamical and electromagnetic contributions,
}
\begin{align}
\phi_{\rm dyn} &= -\frac{mc^2}{\hbar}\int_\gamma d\tau,
\label{eq:dynphase}\\
\phi_{\rm EM} &= \frac{q}{\hbar}\int_\gamma A_\mu\,dx^\mu.
\label{eq:ABphase}
\end{align}
For a closed path the electromagnetic contribution reduces to a
gauge-invariant flux, reflecting its nature as a $U(1)$ holonomy.
{
Throughout this work, $\phi_{\rm EM}$ denotes the general electromagnetic
phase. The canonical field-free Aharonov--Bohm regime is recovered only
when the static magnetic flux is confined to a region inaccessible to the
particle trajectories.}

{Only differences of these phases, accumulated along distinct
trajectories, are observable; matter-wave interferometry thus probes
spacetime geometry and gauge structure through phase differences.}

{
At next-to-leading order in the WKB expansion, the amplitude obeys a
transport equation along the world line. For the charged particle
considered here, the spin evolution contains an electromagnetic
precession generated by the prescribed background field, a
curvature-dependent part generated by the local Lorentz connection,
and---because the guided arms are non-geodesic---acceleration-dependent
(Thomas-precession) terms generated by the guiding force. In this work
we are concerned with the curvature part. The electromagnetic and
Thomas contributions are prescribed by the external guiding and bias
fields: for a shared initial spin state they are common to both
interferometer arms up to proper-time differences, and we assume they
are subtracted as part of the reference evolution. Strictly, this
cancellation holds for equal proper times; the arm proper-time
difference $\Delta\tau$ leaves a residual precession phase of order
$\omega_{\rm prec}\Delta\tau$, suppressed relative to the dynamical
phase $mc^2\Delta\tau/\hbar$ by the ratio
$\hbar\omega_{\rm prec}/(mc^2)\ll1$ and hence negligible in the regime
considered here. Spin precession driven by the GW-induced field
perturbations $\delta\mathbf B$ of Sec.~\ref{sec:EM_response}
constitutes a separate electromagnetic spin channel, which lies outside
the scope of the present analysis. 
}
{
After isolating the
curvature-dependent part of the spin evolution, we adopt the leading
gravitational parallel-transport approximation along the prescribed
interferometer arms and factor the leading WKB amplitude as
}
\begin{equation}
a_0(x) = \mathcal{A}(x)\,\chi(x),
\end{equation}
where $\mathcal{A}(x)$ is a scalar amplitude and $\chi(x)$ is a
normalized spinor satisfying $\bar{\chi}\chi = 1$. The scalar amplitude
evolves according to a conservation law, while
{within this approximation the gravitational part of the}
spinor transport obeys
\begin{equation}\label{eq:transport}
u^\mu (\partial_\mu + \Gamma_\mu)\,\chi = 0.
\end{equation}
{Momentum-space Berry corrections and the associated spin-dependent
deflection of the wave-packet trajectory enter beyond the fixed-path
approximation retained here~\cite{Oancea2023}.}
Its formal solution is the path-ordered exponential
\begin{equation}\label{eq:spintran}
\chi(\tau_2)=U(\tau_2,\tau_1)\chi(\tau_1),
\qquad
U = \mathcal{P}\exp\!\left(-\int_{\tau_1}^{\tau_2}\Gamma_\mu u^\mu\,
d\tau\right),
\end{equation}
which defines the spin parallel transporter associated with the local
Lorentz connection. The operator $U$ takes values in
$\mathrm{Spin}(1,3)$, the double cover of {the proper orthochronous
Lorentz group.}

{
In general, this holonomy is non-Abelian. At first order in the
weak-field expansion adopted throughout this work, however, commutators
of the connection evaluated at different points along the path contribute
only at quadratic order, and the path ordering may be dropped. One may
therefore define the leading-order integrated spin generator
\begin{equation}\label{eq:spinphase}
\Phi_{\rm spin} =
-i\int_{\tau_1}^{\tau_2}
\Gamma_\mu u^\mu\,d\tau ,
\end{equation}
such that
\begin{equation}
U = 1-i\,\Phi_{\rm spin} +\mathcal O(\Gamma^2) = \exp\!\left(-i\,\Phi_{\rm spin}\right)
+\mathcal O(\Gamma^2).
\end{equation}
After projection onto the positive-energy spin subspace in a local rest
frame, the rotational part of $\Phi_{\rm spin}$ is represented by a
Hermitian $SU(2)$ generator.
We emphasize that $\Gamma_\mu$, and hence $\Phi_{\rm spin}$ evaluated
along a single open arm, depends on the choice of local tetrad
$e^\mu{}_a$ (a local Lorentz gauge choice); gauge-invariant observables
arise only from combinations in which both transported spinors are
referred to a common frame, as in Eq.~\eqref{eq:SpinPhase} below.
}
{%
This reduction to an effectively Abelian generator is exact to the
order retained throughout this work and is what underlies the
single-loop square-Mach--Zehnder calculation of
Sec.~\ref{sec:spin_response_square_main}. The full non-Abelian
$\mathrm{Spin}(1,3)$ structure of Eq.~\eqref{eq:spintran} would instead
become dynamically relevant in protocols where the spinor is
transported through connections evaluated at distinct, non-commuting
instants or along non-coplanar paths --- for example, a two-pulse
sequence probing temporally separated segments of the GW waveform, or
a Sagnac-type configuration enclosing spacelike-separated tidal
events, in the spirit of the relativistic spin-gravity treatments of
Refs.~\cite{Borde2000,Javadinezhad2026}. We do not construct such a
protocol here; the present formalism is retained in its
$\mathrm{Spin}(1,3)$ form because it is the structure that emerges
directly from the WKB expansion of the Dirac equation and places the
spin channel on the same covariant footing as the dynamical and
electromagnetic channels, even though only its Abelian limit is
exercised by the specific geometry considered below.
}

Unlike the dynamical and electromagnetic phases, $\Phi_{\rm spin}$ is
matrix-valued; observable spin-dependent phases arise only after
projection onto a chosen spin state. {For a generic spin state, the
interferometric signal may appear as a spin-dependent rotation or
contrast modulation rather than a single scalar phase.}

{At subleading order, the trajectory also receives spin-dependent
corrections of Mathisson--Papapetrou--Dixon type
~\cite{Rudiger1981,Audretsch1981,Hammad2024}. Schematically,}
\begin{equation}\label{eq:MPDlike}
{
\frac{DP^\mu}{d\tau}
=
qF^\mu{}_\nu u^\nu
-\frac{1}{2}
R^\mu{}_{\nu\rho\sigma}
u^\nu \mathcal S^{\rho\sigma}
+\mathcal O(\hbar\,\nabla F,\,\mathcal S^2),
}
\end{equation}
{where $P^\mu$ and $\mathcal S^{\rho\sigma}$ are the momentum and
spin tensor of the localized wave packet, respectively. At leading
order, $P^\mu=\Pi^\mu=mu^\mu$, while spin-dependent corrections can
modify this relation. The first term is the Lorentz force already
contained in the leading-order trajectories, whereas the curvature term
represents spin-dependent backreaction. In the present work we prescribe
the trajectories through the guided-arm approximation and neglect this
backreaction.}
{
The self-consistency of this approximation can be checked
parametrically. The MPD curvature force has magnitude
\begin{equation}\label{eq:MPD_f}
|f_{\rm MPD}|\sim R\,|\mathcal S|\,c\sim\frac{\hbar\,|\ddot h|}{2c},
\end{equation}
using $|\mathcal S|\sim\hbar/2$ and $R\sim|\ddot h|/c^2$.
This should be compared with the force already implicit in holding the particle to
the prescribed tidal trajectory, which for the guided-arm geometry of
Sec.~\ref{sec:interferometric_response} is of order
$m\ddot h L$. Their ratio is
\begin{equation}\label{eq:MPD_frac}
\frac{|f_{\rm MPD}|}{m\ddot h L}
\sim
\frac{\hbar}{2mcL} = \frac{\lambdabar_C}{2L},
\end{equation}
where $\lambdabar_C\equiv\hbar/(mc)$ is the reduced Compton
wavelength---the same Compton-wavelength suppression identified for the
spin phase itself in Sec.~\ref{sec:spin_response_square_main}. At the
level of this parametric force estimate, the MPD backreaction is
therefore smaller than the leading tidal force by
$\lambdabar_C/(2L)\sim10^{-16}$--$10^{-13}$, supporting the guided-arm
approximation for the regime considered here, provided the guiding
mechanism itself does not introduce spin-dependent forces of comparable
or larger size.
}

{
For a two-arm interferometer with paths $\gamma_A$ and $\gamma_B$
recombining at a common detection event, both transported spinors are
projected onto the same positive-energy spin subspace defined by a
common local tetrad at the recombination point. In this local rest-frame
spin space the rotational parts of the transporters are represented by
unitary $SU(2)$ matrices, and the spin-dependent phase shift is
}
\begin{equation}\label{eq:SpinPhase}
\Delta\phi_{\rm spin}
=
\arg\!\left[\langle s|U_B^\dagger U_A|s\rangle\right],
\end{equation}
where $|s\rangle$ denotes the initial spin state
{and, to leading order,
$U_{A,B}=\exp(-i\,\Phi_{\rm spin}[\gamma_{A,B}])+\mathcal O(\Gamma^2)$
with the generator of Eq.~\eqref{eq:spinphase} evaluated along the
corresponding arm.}

{
At the detection event, the leading-order WKB amplitudes superpose as
$\psi_A+\psi_B$ with
$\psi_\gamma=\mathcal A_\gamma U_\gamma|s\rangle\,e^{i\phi_\gamma}$,
yielding the interference signal
\begin{equation}
I\propto
\mathcal A_A^2+\mathcal A_B^2
+2\mathcal A_A\mathcal A_B\,V
\cos\!\left(
\Delta\phi_{\rm dyn}+\Delta\phi_{\rm EM}+\Delta\phi_{\rm spin}
\right),
\label{eq:interference_signal}
\end{equation}
where $V=|\langle s|U_B^\dagger U_A|s\rangle|\le1$ is the spin
visibility and all arm differences are defined with the orientation
$\Delta\phi\equiv\phi_A-\phi_B$. Within the leading-WKB and guided-arm
approximations, the additivity of the three channels follows from the
decomposition in Eq.~\eqref{eq:S_split} together with the factorization
of the amplitude into a real scalar part and a transported spinor.
Channel mixing may arise through spin-dependent corrections to the
trajectories [Eq.~\eqref{eq:MPDlike}], induced electromagnetic spin
precession, residual Thomas-precession imbalance between the arms, or
other higher-order transport effects that are neglected here.
}

The total interferometric phase shift therefore decomposes as
\begin{equation}
\Delta\phi_{\rm tot}
=
\Delta\phi_{\rm dyn} + \Delta\phi_{\rm EM} + \Delta\phi_{\rm spin}.
\end{equation}

\section{Phase structure in a detector frame}
\label{sec:decomp}

The Dirac equation, Eq.\eqref{eq:DiracCurved}, is covariant, whereas the
observable interferometric phase is defined relative to the local frame of
the detector. We therefore evaluate the phase in a detector frame adapted
to a reference world line, where inertial effects (acceleration and
rotation) and spacetime curvature appear as local fields governing the
dynamical phase [Eq.\eqref{eq:dynphase}] and spin phase [Eq.\eqref{eq:spinphase}].
{Our Riemann tensor convention is
\begin{equation}
R^\rho{}_{\sigma\mu\nu} = \partial_\mu\Gamma^\rho_{\nu\sigma} - \partial_\nu\Gamma^\rho_{\mu\sigma} + \Gamma^\rho_{\mu\lambda}\Gamma^\lambda_{\nu\sigma} -
\Gamma^\rho_{\nu\lambda}\Gamma^\lambda_{\mu\sigma},
\end{equation}
where $\Gamma^\rho_{\mu\nu}$ denote the Christoffel symbols, not to be
confused with the spin connection $\Gamma_\mu$ of
Eq.~\eqref{eq:Gamma_mu}.}

Consider a detector following a timelike world line with proper
acceleration $\mathbf a(t)$ and a spatial frame rotating with angular
velocity $\boldsymbol{\Omega}(t)$ relative to a Fermi--Walker transported
frame. We use local coordinates $(x^0,\mathbf x)$ with $x^0=ct${,
adapted to the orthonormal tetrad carried by the detector; the indices
$0,i$ on the curvature tensor below label components in this frame,
evaluated on the reference world line.} For an
apparatus of size $L$ satisfying $L/\mathcal R\ll1$, where
$\mathcal R$ {is the local curvature radius set by the typical
magnitude of the curvature components,
$\mathcal R\sim|R_{\alpha\beta\gamma\delta}|^{-1/2}$,} the curvature is
effectively uniform across the detector
and higher-order tidal corrections are neglected. In this regime, the
metric admits a local expansion about the reference world line in powers
of
$|\mathbf x|$~\cite{MTW,NiZimmermann1978,Marzlin1994,PoissonPoundVega2011}{,
valid within the domain $|\mathbf a\cdot\mathbf x|/c^2\ll1$ and
$|\boldsymbol{\Omega}\times\mathbf x|/c\ll1$.}
{Retaining terms through second order in $|\mathbf x|$, to leading
order in curvature and inertial gradients, while neglecting
$\mathcal O(x^3\nabla R)$ terms, the metric reads}
\begin{subequations}\label{eq:metric_expand}
\begin{align}
g_{00} &= -\left[1 + \frac{2(\mathbf a\cdot\mathbf x)}{c^2}
+\frac{(\mathbf a\cdot\mathbf x)^2}{c^4}
-\frac{(\boldsymbol{\Omega}\times\mathbf x)^2}{c^2} + R_{0i0j}\,x^i x^j\right],
\\
g_{0i}
&= \frac{1}{c}(\boldsymbol{\Omega}\times\mathbf x)_i - \frac{2}{3}R_{0jik}\,x^j x^k,
\\
g_{ij} &= \delta_{ij} - \frac{1}{3}R_{ikjl}\,x^k x^l ,
\end{align}
\end{subequations}
with all curvature components evaluated on the reference world line. For
the gravitational-wave applications considered below, we additionally
assume the long-wavelength limit $L/\lambda_{\rm gw}\ll1$, so that spatial
gradients of the tidal tensors are negligible across the interferometer.

\subsection{Dynamical phase}
\label{sec:dynphase_decomp}

The dynamical phase [Eq.\eqref{eq:dynphase}] is determined by the proper time
accumulated along the particle trajectory{, which in the detector
frame is modulated by both inertial effects and spacetime curvature.}
Let $\mathbf{v}=d\mathbf{x}/dt$. Substituting the metric
\eqref{eq:metric_expand} into
$d\tau^2=-c^{-2}g_{\mu\nu}dx^\mu dx^\nu$ and expanding in the small
quantities introduced above yields
\begin{widetext}
\begin{align}
\frac{d\tau}{dt} &=
1
+ \frac{\mathbf{a}\cdot\mathbf{x}}{c^2}
-\frac{v^2}{2c^2}
- \frac{\mathbf{v}\cdot(\boldsymbol{\Omega}\times\mathbf{x})}{c^2}
- \frac{(\boldsymbol{\Omega}\times\mathbf{x})^2}{2c^2}
{-\frac{v^4}{8c^4}}
+\frac{(\mathbf{a}\cdot\mathbf{x})\,v^2}{2c^4}
+\frac{1}{2}R_{0i0j}x^i x^j
+\frac{2}{3c}R_{0jik}v^i x^j x^k
+\cdots ,
\label{eq:dtaudt_expand_refined}
\end{align}
\end{widetext}
where the omitted terms are of higher order in the detector-frame
expansion{. The two $\mathcal O(c^{-4})$ terms $-v^4/(8c^4)$ and
$(\mathbf a\cdot\mathbf x)v^2/(2c^4)$ arise at the same order in the
expansion of the square root and are retained together for consistency;
$\mathcal O(c^{-4})$ cross terms involving the rotation are neglected
under the assumption $|\boldsymbol{\Omega}\times\mathbf x|\ll v$, and the
term $(\mathbf a\cdot\mathbf x)^2/c^4$ cancels identically at this
order.} Substituting
\eqref{eq:dtaudt_expand_refined} into \eqref{eq:dynphase} gives
\begin{equation}
\phi_{\rm dyn} \simeq \phi_0 + \phi_{\rm kin} + \phi_a
{+ \phi_{v^4}}
+ \phi_{\rm rel} + \phi_{\rm Sag} + \phi_{\rm cent}
+ \phi_{R}^{(E)} + \phi_{R}^{(B)}.
\end{equation}

The rest-mass contribution
\begin{equation}
\phi_0 = -\frac{mc^2}{\hbar}\int dt
\end{equation}
{is common to both arms and cancels in the phase difference.} The
kinetic term
\begin{equation}
\phi_{\rm kin} = \frac{m}{2\hbar}\int v^2\,dt
\end{equation}
reproduces the nonrelativistic phase. The acceleration term
\begin{equation}
\phi_a = -\frac{m}{\hbar}\int (\mathbf a\cdot\mathbf x)\,dt
\end{equation}
{reduces to the Colella--Overhauser--Werner (COW) phase for
$\mathbf a=-\mathbf g$~\cite{colella1975}, consistent with the
equivalence principle.}
{The relativistic corrections comprise the kinetic-energy correction
\begin{equation}
\phi_{v^4}
=
\frac{m}{8\hbar c^2}\int v^4\,dt ,
\end{equation}
familiar from the expansion of the relativistic Lagrangian
$-mc^2/\gamma$, and the cross term}
\begin{equation}
\phi_{\rm rel}
=
-\frac{m}{2\hbar c^2}\int
(\mathbf a\cdot\mathbf x)\,v^2\,dt
{,}
\end{equation}
{which} describes the leading coupling between the inertial potential and the
kinetic energy. Rotation produces the Sagnac phase
\begin{equation}
\phi_{\rm Sag} =
\frac{m}{\hbar}\int
(\boldsymbol{\Omega}\times\mathbf x)\cdot d\mathbf x ,
\end{equation}
which, for a closed loop {traversed while $\boldsymbol{\Omega}$ is
constant (or varies negligibly over the traversal time)}, becomes
\begin{equation}
\Delta\phi_{\rm Sag}
=
\frac{2m}{\hbar}\,\boldsymbol{\Omega}\cdot\mathbf A ,
\end{equation}
{where $\mathbf A$ is the oriented area enclosed by the
interferometer loop}~\cite{Sagnac1913,Werner1979}{, and the
centrifugal contribution}
\begin{equation}
\phi_{\rm cent}
=
\frac{m}{2\hbar}\int
(\boldsymbol{\Omega}\times\mathbf x)^2\,dt
\end{equation}
is subleading relative to the Sagnac term {whenever
$|\boldsymbol{\Omega}\times\mathbf x|\ll v$}.

Spacetime curvature induces the tidal contributions
\begin{align}
\phi_{R}^{(E)}
&=
-\frac{mc^2}{2\hbar}\int R_{0i0j}\,x^i x^j\,dt ,
\label{eq:gravielectricphase}
\\
\phi_{R}^{(B)}
&=
-\frac{2mc}{3\hbar}\int R_{0jik}\,v^i x^j x^k\,dt ,
\end{align}
corresponding to gravitoelectric and gravitomagnetic couplings in the
detector frame. {In flat spacetime these vanish, recovering the
standard inertial phase shifts observed in neutron
interferometry~\cite{Werner1979,varju2000}. Spin-dependent contributions
are treated separately through the spin holonomy discussed next.}

\subsection{{Spin phase}}
\label{sec:spinholonomy_properdetector}

Unlike the dynamical phase, the spin phase is determined by parallel
transport in the {local frame} and is therefore sensitive to
the rotational structure of that frame. The spin connection $\Gamma_\mu$
of Eq.~\eqref{eq:Gamma_mu} generates the transporter [Eq.\eqref{eq:spintran}]
and hence the spin generator [Eq.\eqref{eq:spinphase}]{; its explicit
detector-frame form is derived below.}

To first order in the metric perturbation
$h_{\mu\nu}=g_{\mu\nu}-\eta_{\mu\nu}$, we choose the symmetric tetrad
gauge
\begin{equation}
{e^a{}_\mu = \delta^a{}_\mu + \frac{1}{2}h^a{}_\mu + \mathcal{O}(h^2).}
\end{equation}
{At this order, this gauge realizes the orthonormal frame carried by
the detector introduced in Sec.~\ref{sec:decomp}; it thereby fixes the
local Lorentz gauge on which $\Phi_{\rm spin}$ depends
(Sec.~\ref{sec:quantumphase}) and coincides with the common projection
basis used at the recombination event in Eq.~\eqref{eq:SpinPhase}.}
With the convention $\omega_{\mu ab} = e_a{}^\nu\nabla_\mu e_{b\nu}$, the
spin-connection coefficients reduce at linear order to
\begin{equation}
\omega_{\mu ab} = \frac{1}{2}\left(\partial_b h_{a\mu} - \partial_a
h_{b\mu}\right) + \mathcal{O}(h^2),
\label{eq:omega_linear_detector}
\end{equation}
{where the symmetry $h_{\mu\nu}=h_{\nu\mu}$ has been used.} Using the
detector-frame metric~\eqref{eq:metric_expand} and retaining terms linear
in acceleration, rotation, and curvature, {the connection follows
directly from Eq.\eqref{eq:omega_linear_detector}. Consistent with the
quasi-static treatment of the rotation in
Sec.~\ref{sec:dynphase_decomp}, we take $\boldsymbol{\Omega}$ constant
over the traversal time: the term
$\propto(\dot{\boldsymbol{\Omega}}\times\mathbf x)/c^2$ generated by
$\partial_0 h_{0i}$ enters only the boost sector and is relegated to
$\Phi_{\rm spin}^{(1)}$ below, while terms
$\propto\dot R\,x^2$ lie beyond the order in $|\mathbf x|$ retained
here.}

Introducing the Lorentz generators
\begin{equation}
K^i \equiv \Sigma^{0i},
\qquad
J^i \equiv \frac{1}{2}\epsilon^{ijk}\Sigma^{jk},
\end{equation}
the temporal contribution of the spin connection becomes
\begin{align}
{i\,\Gamma_0\, dx^0}
&=
-\frac{1}{2c}a_iK^i\,dt
-\frac{c}{2}\mathcal E_{ij}x^jK^i\,dt
\nonumber\\
&\quad
-\frac{1}{2}\Omega_iJ^i\,dt
-\frac{c}{2}\mathcal B_{ij}x^jJ^i\,dt ,
\label{eq:Gamma_temporal_detector}
\end{align}
where
\begin{equation}
\mathcal E_{ij} = R_{0i0j},
\qquad
\mathcal B_{ij} = \frac{1}{2}\epsilon_i{}^{kl}R_{0jkl}
\label{eq:EB_tidal_definition}
\end{equation}
{are the detector-frame gravitoelectric and gravitomagnetic tidal
fields, with dimensions
$[\mathcal E_{ij}]=[\mathcal B_{ij}]=L^{-2}$.}

For a vacuum gravitational wave in the TT gauge,
\begin{equation}
\mathcal E_{ij} = -\frac{1}{2c^2}\,\ddot h^{\rm TT}_{ij},
\label{eq:E_TT_detector}
\end{equation}
where dots denote derivatives with respect to coordinate time $t$. For a
plane wave $h^{\rm TT}_{ij}=h^{\rm TT}_{ij}(t-\mathbf n\cdot\mathbf x/c)$,
the linearized vacuum Riemann tensor gives
\begin{equation}
R_{0jkl} = \frac{1}{2c^2}\left(n_k\ddot h^{\rm TT}_{lj}
- n_l\ddot h^{\rm TT}_{kj}\right),
\end{equation}
and hence
\begin{equation}
\mathcal B_{ij} = \frac{1}{2c^2}\,\epsilon_i{}^{kl}\,n_k\,
\ddot h^{\rm TT}_{lj},
\label{eq:B_TT_detector}
\end{equation}
{so that both tensors probe the same local curvature scale,
$\Omega_{\rm gw}^2 h_0/c^2$, where $\Omega_{\rm gw}$ denotes the angular
frequency of the gravitational wave and $h_0$ its characteristic strain
amplitude ($\Omega_{\rm gw}$ is not to be confused with the frame
rotation rate $\boldsymbol{\Omega}$). In the long-wavelength regime the
tidal tensors are evaluated on the reference world line, as stated above,
while their time dependence is retained. The spatial connection
contributes via $dx^i=v^i\,dt$ and is suppressed by $v/c$, entering at
the same order as the boost sector.}

Equation~\eqref{eq:Gamma_temporal_detector} acts on a four-component
Dirac spinor. To extract its nonrelativistic content, we project onto the
positive-energy subspace. Writing
{$\psi=(\phi,\eta)^T$, the small component is determined from the
Dirac equation~\eqref{eq:DiracCurved} at leading order as
\begin{equation}
\eta \simeq \frac{\boldsymbol{\sigma}\cdot\mathbf{p}}{2mc}\,\phi,
\end{equation}
yielding the nonrelativistic scaling $\eta/\phi=\mathcal{O}(v/c)$.}
With the convention of Eq.~\eqref{eq:Gamma_mu}, the rotation generators
act diagonally on the positive-energy component:
{using the standard representation
$\Sigma^{jk} = \epsilon^{jk\ell}\mathrm{diag}(\sigma^\ell,\sigma^\ell)$,}
\begin{equation}
J^i_{\rm eff} = \sigma^i,
\label{eq:J_reduce_corrected}
\end{equation}
whereas the boost generators $K^i=\Sigma^{0i}$ mix the large and small
components, so their contribution is suppressed by one power of $v/c$.
The acceleration and gravitoelectric terms in Eq.\eqref{eq:Gamma_temporal_detector} enter through the boost generators and
are therefore subleading in the nonrelativistic spin holonomy.

{
Substituting Eq.~\eqref{eq:Gamma_temporal_detector} into the spin
generator [Eq.\eqref{eq:spinphase}], and using $\Gamma_\mu u^\mu\,d\tau=\Gamma_\mu\,dx^\mu$ with $dx^0=cdt$ (the difference between $d\tau$ and $dt$ contributes only at higher order),
one has
$\Phi_{\rm spin}=-i\int\Gamma_0\,dx^0+\dots
=-\int(i\,\Gamma_0\,dx^0)+\dots$, and the resulting operator admits an
expansion in powers of $v/c$,
}
\begin{equation}
\Phi_{\rm spin} = \Phi_{\rm spin}^{(0)}
+ \Phi_{\rm spin}^{(1)}
+\cdots,
\label{eq:Phi_spin_expansion}
\end{equation}
where $\Phi_{\rm spin}^{(1)}$ collects the boost-sector and
spatial-connection contributions. At leading order, only the rotation
sector survives:
\begin{equation}
\Phi_{\rm spin}^{(0)} = \frac{1}{2}
\int\boldsymbol{\Omega}\cdot\boldsymbol{\sigma}\,dt
+ \frac{c}{2}\int\mathcal B_{ij}\,x^j\sigma^i\,dt .
\label{eq:Phi_spin_NR0}
\end{equation}
The first term is the Mashhoon spin--rotation coupling~{\cite{Mashhoon1988,HehlNi1990}}; the second is the leading coupling of spin to the gravitomagnetic tidal field, the
spin-sector analogue of $\phi_R^{(B)}$ in the dynamical phase, both
originating from the curvature component $R_{0jik}$.

{The operator $\Phi_{\rm spin}^{(0)}$ generates an $SU(2)$ rotation
in spin space rather than a pure scalar phase, and observable phase
shifts follow after projection onto a specified spin state, as in
Eq.~\eqref{eq:SpinPhase}. For a generic initial state, the spin holonomy
appears as a spin rotation or a visibility modulation. In the
configurations considered below, state preparation and readout in a fixed
spin basis map the leading effect to an effective scalar phase; all
remaining contributions are suppressed by $\mathcal O(v/c)$.}

\section{Electromagnetic response to gravitational waves}
\label{sec:AB_GW}
{
Among the three phase channels identified in Sec.~\ref{sec:quantumphase}, the gravity-induced electromagnetic contribution is qualitatively distinct. Unlike the dynamical and spin phases, which arise directly from the propagation and spin transport of the particle in curved spacetime, this contribution arises indirectly through the response of a
pre-existing electromagnetic field to spacetime curvature. Writing
$A_\mu=A_\mu^{(0)}+\delta A_\mu$, the curvature-induced perturbation
$\delta A_\mu$ produces a corresponding correction to the electromagnetic
phase accumulated along the interferometer paths.
}

Maxwell's equations in curved spacetime are given by
\begin{equation}
\nabla^\nu F_{\mu\nu} = \mu_0 J_\mu,
\end{equation}
where the field strength tensor is defined as $F_{\mu\nu} = \nabla_\mu A_\nu - \nabla_\nu A_\mu$, and SI units are adopted throughout. In the Lorenz gauge, $\nabla_\mu A^\mu = 0$,
these equations reduce to the wave equation:
\begin{equation}
\nabla^\rho\nabla_\rho A_\mu - R_\mu{}^\nu A_\nu = -\mu_0 J_\mu.
\label{eq:maxwell_covariant_full}
\end{equation}
Assuming a vacuum GW background and neglecting the backreaction of the electromagnetic
field and current on the spacetime geometry ($R_{\mu\nu}=0$), Eq.~\eqref{eq:maxwell_covariant_full} reduces to
\begin{equation}
\nabla^\rho\nabla_\rho A_\mu = -\mu_0 J_\mu.
\label{eq:maxwell_vacuum_main}
\end{equation}
{
To determine the resulting electromagnetic response, we analyze the
GW-induced perturbation of a background electromagnetic field in a freely
falling detector frame described by Fermi normal coordinates (FNC). Unlike
the analyses of the dynamical and spin phases, which retained general
detector acceleration and rotation, we neglect these inertial effects to
isolate the GW contribution. We further restrict our attention to the local
detector regime, $L \ll \mathcal{R}=\mathcal{O}(c/(\Omega_{\rm gw}\sqrt{h_0}))$,
where $L$ is the detector size and $\mathcal{R}$ is the curvature radius of
a weak monochromatic GW, ensuring the validity of the FNC expansion.
Additionally, we assume the long-wavelength approximation,
$L\Omega_{\rm gw}/c \ll 1$, or equivalently $L \ll \lambdabar_{\rm gw}=c/\Omega_{\rm gw}$,
so that the curvature is approximately spatially uniform across the detector. Accordingly, terms involving spatial gradients of the curvature in the FNC expansion, such as $x^k\nabla_kR_{\alpha\beta\gamma\delta}$, are neglected, whereas the full temporal dependence of the curvature evaluated on the reference worldline is retained.
}

Retaining terms through $\mathcal{O}(Rx^2)$ in the FNC expansion, the metric perturbation
$h_{\mu\nu}$ can be expressed, for a vacuum GW background,
in terms of the gravitoelectric and gravitomagnetic tidal tensors, defined in Eq.\eqref{eq:EB_tidal_definition}, as
\begin{subequations}
\label{eq:fnc_metric_tidal_main}
\begin{align}
h_{00} &= -\mathcal{E}_{ij}x^ix^j, \\
h_{0i} &= -\frac{2}{3} \epsilon_{ik}{}^l\mathcal{B}_{lj}x^jx^k, \\
h_{ij} &= \frac{1}{3} \epsilon_{ik}{}^m\epsilon_{jl}{}^n \mathcal{E}_{mn}x^kx^l.
\end{align}
\end{subequations}

Expanding Eq.\eqref{eq:maxwell_vacuum_main} to linear order in $h_{\mu\nu}$ yields
a coupled system of equations for the components of $A_\mu$, with the explicit
derivation detailed in Appendix~\ref{app:maxwell_fnc_vacuum_gw}. This expansion takes the form
\begin{widetext}
\begin{equation}
\nabla^\rho\nabla_\rho A_\mu =
\square A_\mu - h^{\alpha\beta}\partial_\alpha\partial_\beta A_\mu
- \mathcal{C}^\lambda\partial_\lambda A_\mu
- \mathcal{D}_\mu{}^{\beta\lambda}\partial_\beta A_\lambda
- \frac{1}{2}\mathcal{P}_\mu{}^\lambda A_\lambda
+ \mathcal{O}(h^2),
\label{eq:maxwell_operator_main}
\end{equation}
\end{widetext}
where $\square = -\frac{1}{c^2}\partial_t^2 + \delta^{ij}\partial_i\partial_j$.
The coefficients $\mathcal{C}^\lambda$, $\mathcal{D}_\mu{}^{\beta\lambda}$, and
$\mathcal{P}_\mu{}^\lambda$ arise from the FNC expansion of the connection
and its derivatives, whereas the term proportional to $h^{\alpha\beta}$
accounts for the correction from the inverse metric.

The component equations derived in Appendix~\ref{app:maxwell_fnc_vacuum_gw}
contain three classes of curvature-induced terms: modifications to the
principal wave operator, derivative couplings among the components of
$A_\mu$, and algebraic couplings constructed from the tidal tensors and
their derivatives. These equations are supplemented by the Lorenz-gauge
constraint, which to linear order in $h_{\mu\nu}$ becomes
\begin{equation}
\partial_\mu A^\mu
- h^{\mu\nu}\partial_\mu A_\nu
- \left( \partial^\alpha h_\alpha{}^\nu
- \frac{1}{2}\partial^\nu h \right) A_\nu = 0,
\label{eq:lorenz_linear_main}
\end{equation}
where $h=\eta^{\mu\nu}h_{\mu\nu}$. Throughout this section, indices are raised
and lowered with the background Minkowski metric.

Within this formulation, spacetime curvature modifies the propagation of the
electromagnetic field through curvature-dependent terms in the wave equation,
which may be interpreted as effective source terms driving electromagnetic
perturbations in the detector frame~\cite{Ruggiero2025}.
{
While detector-frame formulations of Maxwell's equations in gravitational-wave
backgrounds have been developed previously \cite{Marklund2000,Brodin2001,Servin2003,Domcke2022,Domcke2024,Ruggiero2025},
our purpose here is not to introduce a new formulation. Instead, we derive the
local weak-field equations directly from the covariant Maxwell equations within
the specific assumptions adopted throughout this work, namely the Fermi normal
detector frame together with the long-wavelength approximation. This provides a
self-consistent starting point for evaluating the curvature-induced
electromagnetic phase of charged matter-wave interferometers and establishes a
framework that can be extended in future investigations.
}

\subsection{Gravitational-wave modulation of a uniform magnetic field}
\label{sec:EM_response}

The modulation of a uniform magnetic field by a GW induces an
electromagnetic response relevant for the electromagnetic phase. This
response is analyzed in the detector frame using the Maxwell--Lorenz
system derived in Appendix~\ref{app:maxwell_fnc_vacuum_gw}.

Consider a uniform background magnetic field
\begin{equation}
\mathbf B^{(0)} = B_0 \hat{\mathbf z},
\end{equation}
with corresponding vector potential
\begin{equation}
A_x^{(0)} = -\frac12 B_0 y,\quad
A_y^{(0)} = \frac12 B_0 x,\quad
A_z^{(0)} = 0,\quad
A_0^{(0)} = 0 .
\end{equation}
{
The current $J^\mu$ that sources and maintains $\mathbf B^{(0)}$ is not
part of the interferometer region: it is treated as an external,
prescribed current lying outside the domain of the FNC expansion, and is
held fixed in the local detector frame (i.e., unperturbed by the GW) at
the order retained here. Its own curvature-induced response would enter
at the same order as the backreaction terms already dropped in
Eq.~\eqref{eq:maxwell_vacuum_main} and is not considered in this work.
}

We introduce
\begin{equation}
A_\mu=\left(-\frac{\phi}{c},\mathbf A\right),
\qquad F_{0i}=-\frac{E_i}{c},
\qquad F_{ij} = \epsilon_{ijk}B_k .
\end{equation}
The vector potential is expanded as
\begin{equation}
A_\mu=A_\mu^{(0)}+A_\mu^{(1)},
\qquad
A_\mu^{(1)}=\mathcal O(h),
\end{equation}
and only terms linear in the GW amplitude are retained. Accordingly,
curvature-dependent operators acting on the induced field
$A_\mu^{(1)}$ are consistently neglected, since they contribute only at
$\mathcal O(h^2)$.

{
The reduction of the detector-frame equations is controlled by the
long-wavelength condition $L\ll\lambdabar_{\rm gw}=c/\Omega_{\rm gw}$
introduced in Sec.~\ref{sec:AB_GW}. In the FNC expansion, every time
derivative of a tidal tensor enters the field equations accompanied by a
factor of $1/c$ and one additional power of the spatial coordinate, as
seen explicitly in Eqs.~\eqref{eq:C_components_app}--\eqref{eq:Pk_spatial_app}. Relative to
the corresponding undotted terms, such contributions are suppressed by
\begin{equation}
\frac{|\dot{\mathcal E}_{ij}|\,x/c}{|\mathcal E_{ij}|}
\sim \frac{\Omega_{\rm gw}L}{c}
= \frac{L}{\lambdabar_{\rm gw}} \ll 1,
\end{equation}
and similarly for $\dot{\mathcal B}_{ij}$. All terms involving time
derivatives of the tidal tensors are therefore dropped from the field
equations, while the full time dependence of $\mathcal E_{ij}(t)$ and
$\mathcal B_{ij}(t)$, evaluated on the reference world line, is
retained. No additional quasi-static assumption is required at the level
of the field equations.
}

Applying these approximations to the detector-frame equations for $A_0$
and $A_k$, the field equations reduce to
\begin{equation}
\square A_0^{(1)}=S_0,
\qquad
\square A_k^{(1)}=S_k,
\label{eq:reduced_wave_main}
\end{equation}
where $\square=-c^{-2}\partial_t^2+\nabla^2$ is the flat-spacetime
d'Alembertian.
{
Evaluating the curvature-dependent operators on the background potential
$A_\mu^{(0)}$, the sources are linear in the spatial coordinates,
\begin{align}
S_0 &= 2B_0\,\mathcal B_{3j}\,x^j,
\label{eq:S0_main}
\\
S_k
&=
-\frac{B_0}{6}
\left(
\epsilon_{k3m}\mathcal E_{mj}
+
\mathcal E_{km}\epsilon_{m3j}
\right)x^j
-
B_0\,
\epsilon_{kjm}\mathcal E_{m3}\,x^j ,
\label{eq:Sk_main}
\end{align}
where the first two contributions in $S_k$ originate from the
first-derivative coupling $\mathcal C^m\partial_m$ and the algebraic term
$\mathcal P_k{}^i$, while the last term arises from the derivative-mixing
coefficient $\mathcal D_k{}^{ji}$ and dominates the response. The scalar
potential is thus sourced by the gravitomagnetic tidal field, while the
vector potential is driven by the gravitoelectric sector.
}

The Lorenz gauge condition reduces to
\begin{equation}
-\partial_0 A_0^{(1)} + \partial_i A_i^{(1)} = G .
\label{eq:lorenz_reduced_main_text}
\end{equation}
{
In the linearized gauge condition,
Eq.~\eqref{eq:lorenz_gauge_tidal_app}, the contribution
$h^{ij}\partial_iA_j^{(0)}$ vanishes identically, since the symmetric
tensor $h^{ij}$ is contracted with the antisymmetric gradient
$\partial_iA_j^{(0)}=\tfrac12B_0\epsilon_{j3i}$. The constraint is
therefore generated solely by $\mathcal C^iA_i^{(0)}$,
\begin{equation}
G
=
-\frac{B_0}{6}\,
\mathcal E_{ij}\,\epsilon_{i3l}\,x^jx^l ,
\label{eq:G_general_main}
\end{equation}
which couples the scalar and vector potentials and fixes the mixed terms
in the solution.
}

Since the sources are polynomial in the spatial coordinates, the solution
for $A_\mu^{(1)}$ is at most quadratic in $\mathbf x$, and we adopt the
ansatz
\begin{equation}
A_\mu^{(1)}(t,\mathbf x)
=
\sum_{\alpha}
C_\alpha(t)\,P_\alpha(\mathbf x),
\end{equation}
with monomials $P_\alpha(\mathbf x)$ up to quadratic order.
{
For linear monomials the spatial Laplacian vanishes and
Eq.~\eqref{eq:reduced_wave_main} yields
$\ddot C_\alpha(t)=-c^2 s_\alpha(t)$, where $s_\alpha(t)$ is the
corresponding source coefficient. With initial data
$C_\alpha(0)=\dot C_\alpha(0)=0$, the coefficients are given by the
double time integral
\begin{equation}
C_\alpha(t)
=
-c^2
\int_0^{t}dt'
\int_0^{t'}dt''\,
s_\alpha(t''),
\label{eq:double_integral_solution}
\end{equation}
valid for arbitrary time dependence of the tidal tensors within the
long-wavelength regime, while the gauge
condition~\eqref{eq:lorenz_reduced_main_text} fixes the quadratic
coefficients. The closed-form expressions quoted below correspond to the
short-time evaluation of Eq.~\eqref{eq:double_integral_solution}, in
which $\ddot h_A$ varies negligibly over the integration interval; their
range of validity is discussed at the end of this section. The full
derivation is presented in Appendix~\ref{app:solution_perp}.
}

{
Since Eq.~\eqref{eq:reduced_wave_main} is a wave equation, its general
solution is the particular polynomial solution plus homogeneous solutions
of the flat-spacetime wave equation, which represent freely propagating
electromagnetic waves unrelated to the GW. The vanishing initial data
adopted above express the physical requirement that no electromagnetic
radiation is present before the arrival of the wave and thereby select
the purely GW-driven response; this is a physical specification rather
than a mathematical necessity, and any pre-existing radiation would
superpose linearly without modifying the GW-induced part. No spatial
boundary condition, such as an outgoing-radiation condition, is imposed:
the polynomial solutions describe the near-zone response within
$|\mathbf x|\lesssim L\ll\lambdabar_{\rm gw}$, the domain of validity of
the detector-frame expansion, and are not intended to extend to the wave
zone. The quantities independent of the choice of initial data---the
effective sources $(S_0,S_k,G)$, the induced fields and their gradients
on the reference world line, and the scaling law
below---constitute the robust predictions of the mechanism, whereas
absolute field values at a given event acquire meaning only once the
electromagnetic environment of the apparatus is specified. For the
shielded laboratory configurations relevant to matter-wave
interferometry, the vanishing-initial-data choice is the natural one.
}

{
Before turning to explicit configurations, we record the characteristic
scales of the response, which follow from the source structure alone.
Since $\mathcal E\sim\ddot h/c^2$, the induced potential over a region of
size $L$ accumulates as
\begin{equation}
A^{(1)}
\sim
B_0\,\ddot h\,L\,t^2
\sim
B_0\,h\,(\Omega_{\rm gw}t)^2 L ,
\end{equation}
and the induced fields scale as $\delta E\sim B_0\ddot h\,L\,t$ and
$\delta B\sim B_0\ddot h\,t^2$, so that
$\delta B/\delta E=\mathcal O(1/c)$. Evaluating at the light-crossing
time $t\sim L/c$ gives
}
\begin{equation}
\delta E \sim \frac{B_0}{c}\ddot h L^2
\sim cB_0 h_0\left(\frac{\Omega_{\rm gw}L}{c}\right)^2 ,
\label{eq:E_scaling}
\end{equation}
showing that the response is governed by the tidal curvature scale
$\ddot h$ and suppressed by $(L/\lambdabar_{\rm gw})^2$ in the
long-wavelength regime. This scaling is consistent with previous
analyses of GW-induced electromagnetic effects, in which the induced
fields scale as $\delta F\sim F^{(0)}\ddot h/c^2$ in the presence of a
background field~\cite{Marklund2000,Servin2003,Brodin2001}, and with
recent proposals based on Rydberg-atom detectors~\cite{Kanno2025}.

\subsubsection{Parallel configuration (\(\mathbf n \parallel \mathbf B_0\))}

{
Consider a GW propagating along $\hat{\mathbf z}$, with
$\mathcal E_{xx}=-\mathcal E_{yy}=\mathcal E_+$,
$\mathcal E_{xy}=\mathcal E_\times$, and
$\mathcal E_A=-\ddot h_A/(2c^2)$. In this configuration the
gravitomagnetic source vanishes, $\mathcal B_{3j}=0$, and the three
contributions to $S_k$ in Eq.~\eqref{eq:Sk_main} cancel identically,
\begin{equation}
S_0=S_k=0 ,
\end{equation}
so the response is driven entirely by the gauge constraint, which for
this orientation reads
\begin{equation}
G
=
\frac{B_0}{6}
\left[
2\mathcal E_+\,xy
+
\mathcal E_\times\left(y^2-x^2\right)
\right].
\label{eq:G_para_main}
\end{equation}
The solution consistent with Eqs.~\eqref{eq:reduced_wave_main} and
\eqref{eq:lorenz_reduced_main_text} is purely scalar,
\begin{equation}
A_i^{(1)}=0,
\qquad
A_0^{(1)}
=
-\frac{cB_0\,t}{6}
\left[
2\mathcal E_+\,xy
+
\mathcal E_\times\left(y^2-x^2\right)
\right],
\end{equation}
where the harmonic monomials $xy$ and $y^2-x^2$ satisfy
$\nabla^2A_0^{(1)}=0$, consistent with $S_0=0$. The response is purely
electric,
\begin{subequations}
\begin{align}
\delta E_x
&=
\frac{B_0}{6}
\left(
\ddot h_+\,y-\ddot h_\times\,x
\right)t,
\\
\delta E_y
&=
\frac{B_0}{6}
\left(
\ddot h_+\,x+\ddot h_\times\,y
\right)t,
\\
\delta E_z
&=0,
\end{align}
\end{subequations}
with $\delta\mathbf B=\mathbf 0$. The induced field is transverse to
$\mathbf B^{(0)}$, grows linearly in time at short times, and vanishes on
the reference world line, reflecting its tidal origin.
}

\subsubsection{Perpendicular configuration (\(\mathbf n \perp \mathbf B_0\))}

{
Consider a GW propagating along $\mathbf n=\hat{\mathbf x}$, with
$\mathcal E_{yy}=-\mathcal E_{zz}=\mathcal E_+$,
$\mathcal E_{yz}=\mathcal E_\times$, and
$\mathcal E_A=-\ddot h_A/(2c^2)$. In this case both tidal sectors
contribute: the gravitomagnetic components
$\mathcal B_{3j}$ source the scalar potential,
$S_0=-2B_0(\mathcal E_+y+\mathcal E_\times z)$, while the components of
Eq.~\eqref{eq:Sk_main} combine to
\begin{equation}
S_x=\frac{7B_0}{6}\left(\mathcal E_+ y+\mathcal E_\times z\right),\;
S_y=-\frac{7B_0}{6}\mathcal E_+ x,\;S_z=-\frac{7B_0}{6}\mathcal E_\times x ,
\label{eq:Sk_perp_components}
\end{equation}
and the gauge source reduces to
\begin{equation}
G
=
-\frac{B_0}{6}\,x\,
\bigl(\mathcal E_+ y + \mathcal E_\times z\bigr).
\label{eq:G_perp_main}
\end{equation}
Solving the reduced system together with the gauge constraint
(Appendix~\ref{app:solution_perp}) yields
\begin{subequations}
\begin{align}
\delta B_x &= 0,
\\
\delta B_y &=
\frac{7B_0}{12}\ddot h_\times\,t^2
-\frac{B_0}{c}\ddot h_\times\,x t,
\\
\delta B_z &=
-\frac{7B_0}{12}\ddot h_+\,t^2
+\frac{B_0}{c}\ddot h_+\,x t ,
\end{align}
\end{subequations}
and
\begin{subequations}
\begin{align}
\delta E_x
&=
-\frac{2B_0}{3}
\left(
\ddot h_+\,y+\ddot h_\times\,z
\right)t,
\\
\delta E_y
&=
-\frac{cB_0}{2}\ddot h_+\,t^2
+\frac{B_0}{2}\ddot h_+\,x t
+\frac{B_0}{2c}\ddot h_+\!\left(y^2-x^2\right),
\\
\delta E_z
&=
-\frac{cB_0}{2}\ddot h_\times\,t^2
+\frac{B_0}{2}\ddot h_\times\,x t
+\frac{B_0}{2c}\ddot h_\times\!\left(z^2-x^2\right).
\end{align}
\end{subequations}
}

In contrast to the parallel configuration, both electric and magnetic
responses are generated. The magnetic perturbation contains a spatially
uniform component together with a gradient term along the propagation
direction, while the electric field exhibits homogeneous, linear, and
quadratic spatial contributions. {The decomposition into uniform and
gradient terms refers to the FNC origin on the reference world line;
frame-independent statements are provided by the field values and their
gradients on that world line and by the scaling
law~\eqref{eq:E_scaling}.} This anisotropic structure is consistent with
previous studies of GW--electromagnetic coupling, which show that the
response depends sensitively on the relative orientation between the
propagation direction and the background
field~\cite{Servin2003,Brodin2001}.

A key structural feature is the role of the Lorenz gauge constraint. Its
enforcement introduces quadratic spatial contributions to the potentials,
which are required for internal consistency and contribute nontrivially
to the induced fields. Such polynomial spatial dependence is
characteristic of the detector-frame description, in which the metric
expansion is truncated at $\mathcal O(Rx^2)$.

{
\paragraph{Effective sources and Gertsenshtein-type coupling.}
The induced fields do not satisfy the source-free Maxwell equations.
Combining Eqs.~\eqref{eq:reduced_wave_main} and
\eqref{eq:lorenz_reduced_main_text}, one finds
\begin{equation}
\nabla\cdot\delta\mathbf E
=
cS_0-\partial_t G
\equiv
\frac{\rho_{\rm eff}}{\varepsilon_0},
\end{equation}
so that the combination of spacetime curvature and the background
magnetic field acts as an effective charge density of order
$\rho_{\rm eff}\sim\varepsilon_0 B_0\ddot h\,L/c$. This is a near-zone,
long-wavelength analogue of the Gertsenshtein mechanism, in which a GW
propagating through a magnetized region converts into an electromagnetic
disturbance: here the conversion appears not as a propagating photon but
as a local field pattern tied to the tidal tensors. The transduction is
linear in both $B_0$ and $h_{0}$, which is the property exploited by
the interferometric phase in Sec.~\ref{sec:AB_response_square}.
}

\paragraph{{Validity of the polynomial time dependence.}}
{%
No temporal quasi-static approximation is imposed in deriving the
reduced field equations~\eqref{eq:reduced_wave_main}; the tidal tensors
may retain their full time dependence. The explicit polynomial
solutions quoted above introduce an additional short-time
approximation, in which the tidal tensors vary negligibly during the
integration interval. They therefore require
}
\begin{equation}
{\Omega_{\rm gw} t \ll 1.}
\end{equation}

{%
Consequently, the terms proportional to \(t\) and \(t^2\) do not
represent secular growth.
}

{%
To see this explicitly, consider a coefficient driven by
\(s_\alpha(t)=s_{\alpha0}\cos(\Omega_{\rm gw}t)\).
Equation~\eqref{eq:double_integral_solution}, with
\(C_\alpha(0)=\dot{C}_\alpha(0)=0\), gives
}
\begin{equation}
{%
C_\alpha(t)
=
\frac{c^2 s_{\alpha0}}{\Omega_{\rm gw}^2}
\left[
\cos(\Omega_{\rm gw}t)-1
\right].
}
\label{eq:monochromatic_EM_coefficient}
\end{equation}

{%
The response is therefore bounded and oscillatory, with
}
\begin{equation}
{%
\left|C_\alpha(t)\right|
\leq
\frac{2c^2\left|s_{\alpha0}\right|}{\Omega_{\rm gw}^2}.
}
\end{equation}

{%
Its short-time expansion is
}
\begin{equation}
{%
C_\alpha(t)
=
-\frac{c^2 s_{\alpha0}}{2}\,t^2
+
\mathcal{O}\!\left(
c^2 s_{\alpha0}\Omega_{\rm gw}^2 t^4
\right),
}
\end{equation}
{%
which reproduces the polynomial solutions above. For the linear
spatial monomials relevant here,
\(s_{\alpha0}P_\alpha(\mathbf{x})
\sim B_0L\Omega_{\rm gw}^2h_0/c^2\)
over a region of size \(L\). The corresponding bounded contribution
to the vector potential is therefore of order
\(C_\alpha P_\alpha\sim B_0Lh_0\), up to geometry- and
boundary-condition-dependent coefficients.
}

{%
For an interferometer, the relevant interval is the transit time
\(\tau=L/v\). Thus, the polynomial expressions may be used directly
only when \(\Omega_{\rm gw}\tau\ll1\). Although the spatial
long-wavelength condition \(\Omega_{\rm gw}L/c\ll1\) controls the
Fermi-normal-coordinate expansion, it does not by itself imply
\(\Omega_{\rm gw}\tau\ll1\) for a nonrelativistic matter wave. Outside
this temporal regime, the polynomial potentials must be replaced by
the full convolution solutions obtained from
Eq.~\eqref{eq:double_integral_solution}, together with the corresponding
Lorenz-gauge constraint.
}

The resulting perturbed fields \(\delta\mathbf{E}\) and
\(\delta\mathbf{B}\) determine the corresponding perturbation of the
gauge potential, which enters the spacetime holonomy relevant for the
electromagnetic phase. In Sec.~\ref{sec:AB_response_square}, we use the
short-time response to compute the gravitational-wave-induced
electromagnetic phase in a square interferometer.

\section{Interferometric response to gravitational waves}
\label{sec:interferometric_response}

As a concrete example, we consider an idealized square Mach--Zehnder
interferometer (MZI) of side length $L$, lying in the $x$--$y$ plane of a
freely falling detector frame.

{
The validity of the description involves several distinct scale
separations. The WKB approximation requires the de Broglie wavelength to
be much shorter than the characteristic scale over which the background
and the guiding potentials vary. The local detector-frame expansion
requires the apparatus size to satisfy $L\ll\mathcal R$, where
$\mathcal R$ is the local curvature radius introduced in
Sec.~\ref{sec:decomp}. In addition, neglecting spatial variations of the
tidal tensors across the apparatus requires the long-wavelength
condition $\Omega_{\rm gw}L/c\ll1$. To ensure a clear hierarchy in this
spatial-gradient expansion, we impose the fiducial condition
\begin{equation}
\frac{\Omega_{\rm gw}L}{c}\lesssim10^{-2},
\qquad\text{equivalently}\qquad
\frac{L}{\lambda_{\rm gw}}\lesssim\frac{10^{-2}}{2\pi},
\end{equation}
with $\lambda_{\rm gw}\equiv2\pi c/\Omega_{\rm gw}=c/f_{\rm gw}$. For
interferometer sizes $L\sim0.1$--$10\,\mathrm{m}$, this corresponds to
$f_{\rm gw}\lesssim5\times10^{6}$--$5\times10^{4}\,\mathrm{Hz}$,
respectively. In this regime, corrections associated with spatial
gradients of the tidal fields are suppressed by the small parameter
$\Omega_{\rm gw}L/c$.
}

We adopt the guided-arm approximation, in which the particle trajectories
are fixed by the interferometer geometry and traversed at constant speed
$v$. In this description, gravitational-wave effects enter not through
deflection of the paths, but through modulation of the phase accumulated
along them.

For the square configuration, the two arms are parametrized as
\begin{subequations}
\label{eq:trajectories_mzi_main}
\begin{align}
\gamma_1:\ (x_1,y_1) &=
\begin{cases}
(vt,0), & 0\le t\le \tau, \\
(L,v(t-\tau)), & \tau\le t\le 2\tau ,
\end{cases} \\
\gamma_2:\ (x_2,y_2) &=
\begin{cases}
(0,vt), & 0\le t\le \tau, \\
(v(t-\tau),L), & \tau\le t\le 2\tau ,
\end{cases}
\end{align}
\end{subequations}
where $\tau=L/v$ is the traversal time along each segment.

{
We do not commit to a specific physical mechanism for enforcing these
trajectories; as in the COW/Stodolsky tradition, the guided-arm
approximation is used as a formal device that isolates the phase
response from trajectory dynamics. A physical realization could be a
rigid material waveguide, a co-moving optical or magnetic trap, or any
sufficiently stiff confining potential, provided its own back-action on
the phase is either common to both arms and subtracted as part of the
reference evolution (Sec.~\ref{sec:quantumphase}) or bounded by the
parametric estimate of Eqs.~\eqref{eq:MPD_f}--\eqref{eq:MPD_frac}.
For the charged-particle electromagnetic channel of
Sec.~\ref{sec:AB_response_square} in particular, realizing this
approximation in practice requires either guiding stiff enough to
suppress the background Lorentz force well below the target phase
sensitivity, or a confined-flux geometry in which the trajectory does not
sample the background field; see the caveats noted in the Discussion
(Sec.~\ref{sec:discussion}).
}

\subsection{Dynamical phase response}
\label{sec:dyn_response_square_main}

{We first consider the dynamical phase, whose leading
nonrelativistic GW response is governed by the gravitoelectric tidal
field.} In a freely falling, nonrotating detector frame and at leading
nonrelativistic order, the dominant contribution arises from the
gravitoelectric tidal term, Eq.~\eqref{eq:gravielectricphase}; the
gravitomagnetic and spatial-curvature contributions are suppressed,
respectively, by factors of order $v/c$ and $v^2/c^2$.

{
The geometry-dependent response kernels introduced below---quadratic
for the dynamical channel and linear for the spin channel---reflect
the spatial weighting of the corresponding couplings: the
gravitoelectric sector couples to the quadratic monomial $x^ix^j$
[Eq.~\eqref{eq:gravielectricphase}], the gravitomagnetic sector to the linear monomial $x^j$
[Eq.~\eqref{eq:Phi_spin_NR0}], and the interferometric response of each channel is
obtained from the arm difference of these monomials evaluated along
the two paths.
}

Using
$R_{0i0j}(t)=-(1/2c^2)\ddot h^{\rm TT}_{ij}(t)$, the phase difference,
defined as $\Delta\phi_{\rm dyn}^{\rm GW}\equiv\phi_1-\phi_2$, becomes
\begin{equation}
\Delta\phi_{\rm dyn}^{\rm GW}
=
\frac{m}{4\hbar}
\int_{0}^{2\tau}dt\,
\ddot h^{\rm TT}_{ij}(t)\,
Q^{ij}(t),
\label{eq:dyn_response_quadrupole_main}
\end{equation}
where
\begin{equation}
Q^{ij}(t)
\equiv
x_1^i(t)x_1^j(t)
-
x_2^i(t)x_2^j(t).
\end{equation}
{Equation~\eqref{eq:dyn_response_quadrupole_main} is consistent with
established detector-frame and proper-time descriptions of matter-wave
interferometric responses to gravitational fields. Accordingly, the
dynamical channel is not introduced here as a new phase mechanism;
rather, it is recovered within the present WKB Dirac framework and
formulated alongside the electromagnetic and spin channels
\cite{Stodolsky1979,Asenbaum2017,Badurina2025,Schaffrath2025,Gaudio2026}.}

For the square MZI trajectories in
Eq.~\eqref{eq:trajectories_mzi_main}, one finds
\begin{equation}
Q^{xy}(t)=0,
\end{equation}
and
\begin{equation}
Q^{xx}(t)
=
-
Q^{yy}(t)
=
L^2 g(t/\tau),
\end{equation}
with
\begin{equation}
g(u) =
\begin{cases}
u^2, & 0\le u\le 1,\\
2u-u^2, & 1\le u\le 2,\\
0, & \text{otherwise}.
\end{cases}
\label{eq:g_kernel_main}
\end{equation}
Thus
\begin{equation}
\Delta\phi_{\rm dyn}^{\rm GW}
=
\frac{mL^2}{4\hbar}
\int_{0}^{2\tau}dt\,
g(t/\tau)
\left[
\ddot h^{\rm TT}_{xx}(t)
-
\ddot h^{\rm TT}_{yy}(t)
\right].
\label{eq:dyn_response_square_components}
\end{equation}

{For a plane GW described in TT coordinates $\mathbf X$ by
\begin{equation}
h^{\rm TT}_{ij}
\!\left(
t-\frac{\mathbf n\cdot\mathbf X}{c}
\right)
=
\sum_{A=+,\times}
h_A
\!\left(
t-\frac{\mathbf n\cdot\mathbf X}{c}
\right)
e^A_{ij}(\mathbf n,\psi),
\end{equation}
where $e^A_{ij}$ are the polarization tensors and $\psi$ is the
polarization angle, the curvature measured on the central detector
worldline is
\begin{equation}
R_{0i0j}(t)
=
-\frac{1}{2c^2}
\sum_A
\ddot h_A(t)\,e^A_{ij}(\mathbf n,\psi).
\end{equation}
When $\Omega_{\rm gw}L/c\ll1$, spatial variations of the curvature across
the apparatus may be neglected, and the phase difference becomes}
\begin{equation}
\Delta\phi_{\rm dyn}^{\rm GW}
=
\frac{mL^2}{4\hbar}
\sum_A
D_A(\mathbf n,\psi)
\int_{0}^{2\tau}dt\,
g(t/\tau)\,
\ddot h_A(t),
\label{eq:dyn_response_polarized}
\end{equation}
where
\begin{equation}
{
D_A(\mathbf n,\psi)
=
e^A_{xx}(\mathbf n,\psi)
-
e^A_{yy}(\mathbf n,\psi).
}
\label{eq:angular_response_dyn}
\end{equation}
{For example, for propagation along $\hat{\mathbf z}$ with the
plus-polarization axes aligned with the interferometer axes
($\psi=0$ in the convention adopted here), one has $D_+=2$ and
$D_\times=0$.}

{In the low-frequency limit $\Omega_{\rm gw}\tau\ll1$, the curvature
varies only weakly during one traversal, and the integral may be
expanded in powers of $\Omega_{\rm gw}\tau$. At leading order,
$\ddot h_A(t)\simeq\ddot h_A(t_*)$, where $t_*$ is a representative time
within the traversal interval.} Since
\begin{equation}
\int_0^{2\tau} g(t/\tau)\,dt = \tau,
\end{equation}
one obtains
\begin{equation}
\Delta\phi_{\rm dyn}^{\rm GW}
\simeq
\frac{mL^3}{4\hbar v}
\sum_A
D_A(\mathbf n,\psi)\,
\ddot h_A(t_*).
\label{eq:dyn_response_lowfreq}
\end{equation}

{This temporal adiabaticity condition is distinct from the spatial
long-wavelength condition. Indeed,
\begin{equation}
\Omega_{\rm gw}\tau
=
\frac{c}{v}
\left(
\frac{\Omega_{\rm gw}L}{c}
\right)
=
2\pi\,\frac{L}{\lambda_{\rm gw}}\,\frac{c}{v},
\label{eq:temporal_vs_spatial}
\end{equation}
so, because $v\ll c$ for nonrelativistic matter waves, the curvature may
be spatially uniform across the apparatus while still varying appreciably
during a single traversal. This distinction applies equally to the spin
and electromagnetic channels below.}

For a monochromatic GW with polarization amplitudes $h_{A0}$ and angular
frequency $\Omega_{\rm gw}$,
{$|\ddot h_A|\sim\Omega_{\rm gw}^2|h_{A0}|$, and the phase amplitude
scales as
\begin{equation}
\left|\Delta\phi_{\rm dyn}^{\rm GW}\right|
\sim
\frac{mL^3\Omega_{\rm gw}^2}{4\hbar v}
\left|
\sum_A
D_A(\mathbf n,\psi)\,
h_{A0}e^{i\varphi_A}
\right|,
\end{equation}
where $\varphi_A$ denotes the phase of each polarization component.
Suppressing numerical, polarization, and angular-response factors, this
reduces to}
\begin{equation}
\left|\Delta\phi_{\rm dyn}^{\rm GW}\right|
\sim
\frac{mL^3}{\hbar v}
\Omega_{\rm gw}^2 h_0 .
\end{equation}
This scaling reflects the tidal nature of the response: the
interferometer is sensitive to the local curvature, encoded in
$\ddot h_A$, rather than directly to the metric perturbation.

\subsection{Spin phase response}
\label{sec:spin_response_square_main}

The spin holonomy provides a complementary channel governed by the
gravitomagnetic sector of the curvature. In the nonrelativistic limit,
the leading contribution is generated by the gravitomagnetic tidal field
$\mathcal B_{ij}$. The observable spin phase is
\begin{equation}
\Delta\phi_{\rm spin}
=
\arg\!\left[\langle s|U_2^\dagger U_1|s\rangle\right],
\end{equation}
where {$U_\alpha=\exp(-i\Phi_\alpha)+\mathcal O(\Gamma^2)$} is the
spin transporter along arm $\gamma_\alpha$
{[Eq.~\eqref{eq:spinphase}]}, with $\alpha=1,2$.
{After projection onto the positive-energy spin subspace, the
observable phase is determined by the effective $SU(2)$ generator, and
non-Abelian ordering effects enter only at $\mathcal O(\Gamma^2)$
[cf.~Sec.~\ref{sec:quantumphase}].}

To leading order, the spin generator along an arm is
\begin{equation}
\Phi_\alpha = \frac{c}{2}\int_{\gamma_\alpha} dt\mathcal B_{ij}(t) x_\alpha^j(t)\sigma^i .
\label{eq:spin_generator_arm_main}
\end{equation}
{
For the square-MZI trajectories of Eq.~\eqref{eq:trajectories_mzi_main}, the arm difference of the position is $x^j_1(t)-x^j_2(t)=L\,f(t/\tau)\,\bigl(\delta^j{}_x-\delta^j{}_y\bigr),$
with the piecewise-linear kernel $f$ defined in Eq.~\eqref{eq:f_kernel_spin_main} below.
Substituting this into Eq.~\eqref{eq:spin_generator_arm_main}, the \emph{second} (spatial) index of $\mathcal B_{ij}$ is contracted with the trajectory difference,
 while the \emph{first} index remains contracted with the Pauli
 matrices $\sigma^i$, so that
$\mathcal B_{ij}\bigl(x^j_1-x^j_2\bigr)\sigma^i
=L\,f(t/\tau)\bigl[\mathcal B_{ix}(t)-\mathcal B_{iy}(t)\bigr]\sigma^i$,
 and one finds}
\begin{equation}
\Delta\Phi_{\rm spin}\equiv\Phi_1-\Phi_2
=
\frac{cL}{2}
\int_0^{2\tau}dt\,
f(t/\tau)
\left[
\mathcal B_{ix}(t)
-
\mathcal B_{iy}(t)
\right]\sigma^i ,
\label{eq:spin_response_B_components_main}
\end{equation}
where the linear kernel is
\begin{equation}
f(u)=
\begin{cases}
u, & 0\le u\le 1,\\
2-u, & 1\le u\le 2,\\
0, & \text{otherwise},
\end{cases}
\label{eq:f_kernel_spin_main}
\end{equation}
the spin analogue of the quadratic kernel $g(u)$ of the dynamical
response, Eq.~\eqref{eq:g_kernel_main}.

{Using the plane-wave decomposition in TT coordinates and the
long-wavelength replacement $h_A\to h_A(t)$ of
Sec.~\ref{sec:dyn_response_square_main}, the gravitomagnetic tidal
tensor on the central detector world line is}
\begin{equation}
\mathcal B_{ij}(t)
=
\frac{1}{2c^2}
\sum_A
\epsilon_i{}^{kl}
n_k\,
e^A_{lj}(\mathbf n,\psi)\,
\ddot h_A(t),
\label{eq:Bij_GW_main}
\end{equation}
in agreement with Eq.~\eqref{eq:B_TT_detector}. Substitution into
Eq.~\eqref{eq:spin_response_B_components_main} gives
\begin{equation}
\Delta\Phi_{\rm spin}
=
\frac{L}{4c}
\sum_A
\int_0^{2\tau}dt\,
f(t/\tau)\,
\Xi_{iA}(\mathbf n,\psi)\,
\ddot h_A(t)\,
\sigma^i ,
\label{eq:spin_response_polarized_main}
\end{equation}
with the angular response
\begin{equation}
\Xi_{iA}(\mathbf n,\psi)
=
\epsilon_i{}^{kl}n_k
\left[
e^A_{lx}(\mathbf n,\psi)
-
e^A_{ly}(\mathbf n,\psi)
\right].
\label{eq:Xi_spin_main}
\end{equation}

{In the low-frequency limit $\Omega_{\rm gw}\tau\ll1$
[cf.~Eq.~\eqref{eq:temporal_vs_spatial}], the curvature varies only
weakly during one traversal, and at leading order
$\ddot h_A(t)\simeq\ddot h_A(t_*)$, with $t_*$ a representative time
within the traversal interval.} Since
\begin{equation}
\int_0^{2\tau} f(t/\tau)\,dt
=
\tau,
\end{equation}
one obtains
\begin{equation}
\Delta\Phi_{\rm spin}
\simeq
\frac{L^2}{4vc}
\sum_A
\Xi_{iA}(\mathbf n,\psi)\,
\ddot h_A(t_*)\,
\sigma^i
\equiv
\frac{L^2}{4vc}\,
W_i(t_*)\,\sigma^i .
\label{eq:spin_response_lowfreq_main}
\end{equation}

Writing $\mathbf W=|\mathbf W|\hat{\mathbf u}$, a spin eigenstate
satisfying
$\hat{\mathbf u}\cdot\boldsymbol{\sigma}|s_\pm\rangle=\pm|s_\pm\rangle$
acquires the scalar phase
\begin{equation}
\Delta\phi_{\rm spin}^{(\pm)}
\simeq
{\mp}\,
\frac{L^2}{4vc}\,
|\mathbf W| ,
\label{eq:spin_scalar_phase_pm_main}
\end{equation}
{where the sign follows from
$U_2^\dagger U_1=e^{-i\Delta\Phi_{\rm spin}}+\mathcal O(\Phi^2)$ and the
orientation convention $\Delta\phi\equiv\phi_1-\phi_2$; the magnitude is
convention independent.}

{For a monochromatic GW with polarization amplitudes $h_{A0}$,
$|\ddot h_A|\sim\Omega_{\rm gw}^2|h_{A0}|$, so that
$|\mathbf W|\sim|\Xi|\,\Omega_{\rm gw}^2h_0$ and, suppressing numerical
and angular-response factors,}
\begin{equation}
\left|\Delta\phi_{\rm spin}\right|
\sim
\frac{L^2}{vc}\,
\Omega_{\rm gw}^2 h_0 .
\label{eq:spin_scaling_main}
\end{equation}
{Comparing with the corresponding dynamical scaling of
Sec.~\ref{sec:dyn_response_square_main},} one finds
\begin{equation}
\frac{
\left|\Delta\phi_{\rm spin}\right|
}{
\left|\Delta\phi_{\rm dyn}^{\rm GW}\right|
}
\sim
\frac{\hbar}{mcL}
=
\frac{\lambda_C}{L},
\label{eq:spin_dyn_ratio_main}
\end{equation}
where $\lambda_C=\hbar/(mc)$ is the reduced Compton wavelength. The spin
channel therefore probes the same tidal curvature scale as the dynamical
channel, but is suppressed by the quantum-relativistic factor
$\lambda_C/L$.

\subsection{{Electromagnetic phase response}}
\label{sec:AB_response_square}

The electromagnetic channel provides a complementary probe in which the GW
couples indirectly through the electromagnetic field. Unlike the dynamical
and spin phases, which probe spacetime curvature directly through particle
motion and spin transport, the electromagnetic phase arises from the
curvature-induced modification of the gauge field derived in
Sec.~\ref{sec:EM_response}.

{
Separating the background and induced potentials,
$A_\mu=A_\mu^{(0)}+\delta A_\mu$, the background contributes only the
static flux phase, which is common to repeated measurements and carries
no GW signal. The GW-induced phase is the spacetime holonomy of the
perturbation,
\begin{equation}
\Delta\phi_{\rm EM}^{\rm GW} = \frac{q}{\hbar}
\oint_{\mathcal C} \delta A_\mu\, dx^\mu ,
\label{eq:holonomy_deltaA}
\end{equation}
where $\mathcal C=\gamma_1-\gamma_2$ is the closed spacetime contour
formed by the two arms.
}
By the four-dimensional Stokes theorem,
\begin{equation}
\Delta\phi_{\rm EM}^{\rm GW} = \frac{q}{2\hbar}
\int_\Sigma \delta F_{\mu\nu}\,d\Sigma^{\mu\nu},
\label{eq:dphiU1_def}
\end{equation}
with $\Sigma$ any worldsheet bounded by $\mathcal C$, which renders the
phase manifestly gauge invariant. For time-dependent fields the
observable is this spacetime holonomy rather than a purely spatial
flux~\cite{Wakamatsu2025,Rai2019}.

{
Evaluating Eq.~\eqref{eq:holonomy_deltaA} along the guided trajectories
of Eq.~\eqref{eq:trajectories_mzi_main}, with $dx^0=c\,dt$, gives
\begin{equation}
\begin{aligned}
\Delta\phi_{\rm EM}^{\rm GW}
&=
\frac{q}{\hbar}
\int_0^{2\tau}\!dt\,
\Bigl[c\Delta(\delta A_0)\;+
\\ 
&\qquad
\dot{\mathbf x}_1\!\cdot\!
\delta\mathbf A(t,\mathbf x_1)
-\dot{\mathbf x}_2\!\cdot\!
\delta\mathbf A(t,\mathbf x_2)
\Bigr].
\end{aligned}
\label{eq:U1_line_integral}
\end{equation}
where
$\Delta(\delta A_0)=\delta A_0(t,\mathbf x_1)-\delta A_0(t,\mathbf x_2)$.
For the solutions considered below, the leading contribution arises from
the scalar-potential term, while the vector-potential contribution is
either absent or subleading in the nonrelativistic expansion; in field
language, the response is controlled primarily by the induced electric
field. Because the induced fields are inhomogeneous across the
apparatus, the phase depends on field gradients as well as on the
enclosed flux, in contrast to the static Aharonov--Bohm effect.
}

{
\paragraph{Parallel configuration.}
Within the long-wavelength solution derived in
Sec.~\ref{sec:EM_response}, the parallel configuration
$\mathbf n\parallel\mathbf B_0$ has a purely scalar response,
$\delta A_i=0$, with
$\delta A_0=(B_0t/12c)[\,2\ddot h_+\,xy+\ddot h_\times(y^2-x^2)\,]$.
Along the square trajectories, the arm difference of $xy$ vanishes
identically, while $\Delta(y^2-x^2)=-2L^2g(t/\tau)$, with $g$ the
quadrupole kernel of Eq.~\eqref{eq:g_kernel_main}. Hence only the cross
polarization contributes,
\begin{equation}
\Delta\phi_{\rm EM}^{\rm GW,\parallel}
=
-\frac{qB_0L^2}{6\hbar}
\int_0^{2\tau}\!dt\;
t\,g(t/\tau)\,
\ddot h_\times(t).
\label{eq:U1_parallel_exact}
\end{equation}
In the low-frequency limit $\Omega_{\rm gw}\tau\ll1$, using
$\int_0^{2\tau}t\,g(t/\tau)\,dt=\tfrac{7}{6}\tau^2$ and
$\ddot h_\times(t)\simeq\ddot h_\times(t_*)$, with $t_*$ a
representative time within the traversal interval
[cf.~Sec.~\ref{sec:dyn_response_square_main}],
\begin{subequations}
    \begin{align}\label{eq:U1_parallel_scaling}
        \Delta\phi_{\rm EM}^{\rm GW,\parallel} &\simeq -\frac{7qB_0L^2\tau^2}{36\hbar}\,
\ddot h_\times(t_*),\\
\left|\Delta\phi_{\rm EM}^{\rm GW,\parallel}\right|&\sim\frac{qB_0L^2}{\hbar}\,
h_0\left(\Omega_{\rm gw}\tau\right)^2 ,
    \end{align}
\end{subequations}
where $h_0$ denotes the amplitude of the relevant polarization.
}

{
\paragraph{Perpendicular configuration.}
For the perpendicular configuration $\mathbf n=\hat{\mathbf x}$, the
leading scalar-potential term in the plane $z=0$ is
$\delta A_0=-(B_0t^2/2)\,\ddot h_+\,y+\mathcal O(xy\,t/c)$. The arm
difference $\Delta y=-Lf(t/\tau)$ involves the linear kernel $f$ of
Eq.~\eqref{eq:f_kernel_spin_main}, giving
\begin{equation}
\Delta\phi_{\rm EM}^{\rm GW,\perp}
=
\frac{qcB_0L}{2\hbar}
\int_0^{2\tau}\!dt\;
t^2 f(t/\tau)\,
\ddot h_+(t) + \mathcal O\!\left(\frac{v}{c}\right).
\label{eq:U1_perp_exact}
\end{equation}
With $\int_0^{2\tau}t^2f(t/\tau)\,dt=\tfrac{7}{6}\tau^3$, the
low-frequency limit is
\begin{subequations}
    \begin{align}\label{eq:U1_perp_scaling}
        \Delta\phi_{\rm EM}^{\rm GW,\perp} &\simeq\frac{7qcB_0L\tau^3}{12\hbar}\,
\ddot h_+(t_*),\\
\left|\Delta\phi_{\rm EM}^{\rm GW,\perp}\right| &\sim\frac{qB_0L^2}{\hbar}\,
h_0\left(\Omega_{\rm gw}\tau\right)^2\frac{c}{v}.
    \end{align}
\end{subequations}
}

{
The explicit factors of $t$ and $t^2$ in these expressions originate
from the initial-value solution of Sec.~\ref{sec:EM_response}, so the
time coordinate is measured from the reference time at which the
initial conditions are imposed; here the interferometer cycle is taken
to begin at that reference time. These formulas apply only within the
temporal regime in which that perturbative solution remains valid and
do not describe a stationary response at arbitrarily long observation
times.
}

{
Two features of the results are worth noting. First, within the present
initial-value solution and square geometry, the perpendicular response
carries an additional factor $c/v$ relative to the parallel scaling;
this factor originates from the different spacetime dependence of the
induced scalar potentials in the two configurations, and is substantial
for nonrelativistic matter waves. Second, for the propagation
directions, polarization basis, and interferometer orientation adopted
here, the leading parallel response selects $h_\times$, whereas the
leading perpendicular response selects $h_+$; a rotation of the
polarization basis mixes the two.
}

{
\paragraph{Low-frequency scaling and flux dependence.}
In the regime $\Omega_{\rm gw}\tau\ll1$
[cf.~Eq.~\eqref{eq:temporal_vs_spatial}], the two configurations may be
summarized parametrically as
\begin{equation}
\left|\Delta\phi_{\rm EM}^{\rm GW}\right|
\sim\frac{q\,\Phi_B^{(0)}}{\hbar}\,
h_0\left(\Omega_{\rm gw}\tau\right)^2
\kappa,
\quad
\kappa=
\begin{cases}
1, & \mathbf n\parallel\mathbf B_0,\\[2pt]
c/v, & \mathbf n\perp\mathbf B_0,
\end{cases}
\label{eq:U1_frequency_response}
\end{equation}
where $\Phi_B^{(0)}=B_0L^2$ for the spatially uniform background field
assumed in Sec.~\ref{sec:EM_response}. Outside the low-frequency
regime, the full time integrals in
Eqs.~\eqref{eq:U1_parallel_exact} and \eqref{eq:U1_perp_exact} must be
retained; their value depends on the GW phase and on the initial
conditions used to construct the induced field. If the applied flux is
confined to an area $\mathcal A<L^2$, the solutions derived for a
uniform background no longer apply directly, and the replacement
$\Phi_B^{(0)}\to B_0\mathcal A$ should be regarded as an
order-of-magnitude estimate pending a solution with the appropriate
electromagnetic boundary conditions.
}

{Within the common low-frequency regime $\Omega_{\rm gw}\tau\ll1$,
the three channels scale as}
\begin{equation}
\left|\Delta\phi_{\rm dyn}\right|
\sim
\frac{m L^3}{\hbar v}\,
\Omega_{\rm gw}^2 h_0,
\qquad
\left|\Delta\phi_{\rm spin}\right|
\sim
\frac{L^2}{v c}\,
\Omega_{\rm gw}^2 h_0,
\end{equation}
together with Eq.~\eqref{eq:U1_frequency_response}: all three probe the
same curvature scale $\ddot h\sim\Omega_{\rm gw}^2h_0$ but differ in
their coupling mechanisms---tidal motion, spin--curvature coupling, and
curvature-induced electromagnetic response. {Unlike the dynamical
and spin channels, whose amplitudes are fixed by the particle mass and
spin, the electromagnetic response depends on the externally applied
field, the flux geometry, and the electromagnetic boundary conditions,
and is therefore externally tunable.}

\subsection{Discussion}
\label{sec:discussion}

The three phase channels derived above provide complementary probes of the
same underlying spacetime curvature. In all cases, the response is governed
by the tidal scale
$\ddot h_{+,\times}\sim\Omega_{\rm gw}^2h_0$
and filtered by geometry-dependent kernels set by the interferometer.

The dynamical phase probes the gravitoelectric sector through proper-time
differences,
\begin{equation}
\left|\Delta\phi_{\rm dyn}\right|\sim\frac{mL^3}{\hbar v}\,
\Omega_{\rm gw}^2h_0,
\end{equation}
and constitutes the leading semiclassical response, with its mass and size
dependence reflecting its inertial origin. The spin phase probes the
gravitomagnetic sector through spin--curvature coupling,
\begin{equation}
\left|\Delta\phi_{\rm spin}\right|\sim\frac{L^2}{vc}\,
\Omega_{\rm gw}^2h_0,\quad
\frac{\left|\Delta\phi_{\rm spin}\right|}
     {\left|\Delta\phi_{\rm dyn}\right|} \sim \frac{{\lambdabar_C}}{L} \ll 1,
\end{equation}
{where $\lambdabar_C=\hbar/(mc)$ is the reduced Compton wavelength
introduced in Sec.~\ref{sec:quantumphase}.} This
hierarchy reflects the intrinsically quantum-relativistic nature of the
spin channel and mirrors the gravitoelectric--gravitomagnetic
decomposition of the detector frame~\cite{Ruggiero2022}. {
The spin contribution is thus not merely small but unobservable for any
conceivable apparatus, and should be regarded as a conceptual
consequence of the unified formalism --- establishing that
the gravitomagnetic sector couples to spin transport at all within this
framework --- rather than a candidate detection channel.
}

The electromagnetic channel is qualitatively different: curvature first
perturbs the background field through the detector-frame Maxwell
equations, and the induced gauge field generates a spacetime holonomy,
which is gauge invariant by Eq.~\eqref{eq:dphiU1_def}.

{%
From Sec.~\ref{sec:AB_response_square}, in the low-frequency regime
$\Omega_{\rm gw}\tau\ll1$, the response takes the form
}
\begin{equation}
{%
\left|\Delta\phi_{\rm EM}\right|
\sim
\frac{q\Phi_B^{(0)}}{\hbar}\,
h_0\left(\Omega_{\rm gw}\tau\right)^2\kappa,
\qquad
\tau=\frac{L}{v},
\label{eq:EM_scaling_discussion}
}
\end{equation}
{%
where $\Phi_B^{(0)}=B_0L^2$ for the uniform background field of
Sec.~\ref{sec:EM_response}, and $\kappa=1$ and $c/v$ for propagation
parallel and perpendicular to $\mathbf B^{(0)}$, respectively. Outside
this regime, the polynomial potentials must be replaced by the full
convolution solutions of Eq.~\eqref{eq:double_integral_solution},
together with the corresponding Lorenz-gauge constraint, before the
interferometric phase is evaluated.
}

{%
Three features are worth emphasizing. First, the amplitude is set by the
applied flux rather than by the particle mass or spin. Second, within the
present initial-value solution and square geometry, the perpendicular
response carries an additional factor $c/v$, originating from the
different spacetime dependence of the induced scalar potentials in the
two configurations; for nonrelativistic matter waves, this factor is
large. Third, for the polarization basis and interferometer orientation
adopted here, the parallel and perpendicular configurations select
$h_\times$ and $h_+$, respectively, so that the orientation of
$\mathbf B^{(0)}$ acts as a polarization filter for this geometry. Since
the induced fields are inhomogeneous across the apparatus, the phase
depends on field gradients as well as on the enclosed flux, in contrast
to the static Aharonov--Bohm effect. The electric contribution dominates
in the nonrelativistic regime, consistent with previous
analyses~\cite{Marklund2000,Servin2003}.
}

{%
To illustrate the relative magnitudes of the three channels,
consider the representative parameters
}
\begin{equation}
{%
m=m_e,
\; L=1\,{\rm m}, \; v=10^3\,{\rm m\,s^{-1}},
\; f_{\rm gw}=50\,{\rm Hz},\;
h_0=10^{-21}.
}
\end{equation}
{%
The traversal time is $\tau=L/v=10^{-3}\,{\rm s}$, so that
$\Omega_{\rm gw}\tau\simeq0.31$, placing the example marginally within
the low-frequency regime; corrections to the low-frequency kernels are
of order $(\Omega_{\rm gw}\tau)^2\simeq10\%$ and immaterial at the
order-of-magnitude level. The dynamical and spin scalings then give
}
\begin{align}
{\left|\Delta\phi_{\rm dyn}\right|}
&{\sim 9\times10^{-16},}
\\
{\left|\Delta\phi_{\rm spin}\right|}
&{\sim 3\times10^{-28}.}
\end{align}
{%
Their ratio is
}
\begin{equation}
{%
\frac{\left|\Delta\phi_{\rm spin}\right|}
     {\left|\Delta\phi_{\rm dyn}\right|}
\sim
4\times10^{-13},
}
\end{equation}
{%
consistent with $\lambdabar_C/L$. The dynamical result is consistent with
the established tidal matter-wave response demonstrated experimentally
and developed in previous theoretical
analyses~\cite{Asenbaum2017,Badurina2025,Schaffrath2025,Gaudio2026}.
}

{%
Evaluating the electromagnetic scaling
\eqref{eq:EM_scaling_discussion} for the same parameters with a uniform
background field $B_0=1\,{\rm T}$ gives
$qB_0L^2/\hbar\simeq1.5\times10^{15}$ and hence
$|\Delta\phi_{\rm EM}|\sim1.5\times10^{-7}$ in the parallel
configuration and $\sim5\times10^{-2}$ in the perpendicular one,
exceeding the dynamical channel by roughly eight and thirteen orders of
magnitude, respectively. If the applied flux is instead confined
to an area $\mathcal A\ll L^2$, the formal replacement
$B_0L^2\to B_0\mathcal A$ suggests the corresponding order-of-magnitude
scaling; the parallel response then becomes comparable to the dynamical
channel for $B_0\mathcal A\sim10^{-8}\,{\rm T\,m^2}$. This replacement
is not a prediction of the uniform-field
solution, however, because the confined-flux configuration requires
different electromagnetic boundary conditions. We therefore use it only
to illustrate the possible flux dependence and not as a quantitative
detector estimate.
}

{%
These estimates must be interpreted with care. They follow from the
low-frequency scaling laws and are not sensitivities of a realizable
detector. In particular, the guided-arm approximation treats the
trajectories as fixed, whereas, for charged particles, the background
Lorentz force would generally alter the trajectories unless the arms are
rigidly guided or the flux is confined away from the paths. The formal
extrapolation to a confined flux $\mathcal A\ll L^2$ lies precisely in
the regime in which the uniform-field solution of
Sec.~\ref{sec:EM_response} no longer applies directly. Coherence times
exceeding the interrogation time, together with control of stray fields,
patch potentials, magnetic-field gradients, and collisional dephasing,
would further constrain any implementation. The numerical values above
therefore illustrate the hierarchy among the three channels and the
parametric dependence of the electromagnetic response on the applied
flux; none should be interpreted as a projected detector sensitivity.
}

{%
A practical implementation would also require adequate coherence and
phase sensitivity. For the square trajectories considered here, each
interferometer arm consists of two segments of length $L$, so the
interrogation time for a particle propagating at speed $v$ is
}
\begin{equation}
{%
T_{\rm int}=2\tau=\frac{2L}{v},
}
\end{equation}
{%
giving $T_{\rm int}=2\,{\rm ms}$ for the representative values above.
The coherence time should therefore be at
least comparable to $T_{\rm int}$ and preferably substantially longer
to avoid a significant loss of fringe visibility.
}

{%
Even if coherence is maintained, particle shot noise presents a severe
statistical limitation. For $N$ independent detected particles, fringe
visibility $\mathcal{V}$, and operation near the point of maximum phase
sensitivity, the ideal shot-noise-limited phase uncertainty is
}
\begin{equation}
{%
\delta\phi_{\rm SN}
\simeq
\frac{1}{\mathcal{V}\sqrt{N}}.
}
\end{equation}
{%
Resolving a phase $|\Delta\phi|$ with a signal-to-noise ratio of order
unity therefore requires
}
\begin{equation}
{%
N
\gtrsim
\frac{1}{\mathcal{V}^{2}|\Delta\phi|^{2}}.
}
\end{equation}
{%
For unit visibility, the representative values
$|\Delta\phi_{\rm dyn}|\simeq 9\times10^{-16}$ and
$|\Delta\phi_{\rm spin}|\simeq 3\times10^{-28}$ would require
approximately
$N_{\rm dyn}\gtrsim 10^{30}$ and
$N_{\rm spin}\gtrsim 10^{55}$, respectively --- far beyond realistic
values.
}

{%
For charged particles, stray electric and magnetic fields, patch
potentials, field gradients, Lorentz-force-induced trajectory changes,
collisions, and fluctuations of the applied magnetic field would
introduce additional phase shifts, noise, and dephasing. A realistic
experiment would require the fluctuations and residual uncertainties
associated with these effects, particularly those overlapping the
signal bandwidth, to remain below the gravitational-wave-induced
signal. A quantitative noise budget is apparatus dependent and lies
beyond the scope of the present theoretical analysis.
}

\section{Conclusion}
\label{sec:conclusion}

A matter wave propagating through a GW spacetime accumulates phase that
encodes the underlying geometry. We have developed a semiclassical
framework that makes this imprint explicit for charged
spin-$\tfrac{1}{2}$ matter-wave interferometers. Starting from a WKB
expansion of the covariant Dirac equation, the interferometric phase
separates, at leading order and within the guided-arm approximation,
into three additive contributions: a dynamical phase associated with
proper-time evolution, a spin phase arising from parallel transport in
the local Lorentz frame, and an electromagnetic phase generated by
curvature-induced perturbations of a background electromagnetic field.
The corresponding observables are the proper-time difference, the
projected $SU(2)$ spin holonomy, and the spacetime electromagnetic
holonomy of the induced gauge field, respectively; each is gauge
invariant.

In a {freely falling, non-rotating} detector frame described by
Fermi normal coordinates{---the specialization $\mathbf a=0$,
$\boldsymbol{\Omega}=0$ of the general accelerated, rotating frame of
Sec.~\ref{sec:decomp}, which the framework equally
accommodates---}all responses are governed by local tidal fields:
{at leading order,} the
dynamical phase probes the gravitoelectric sector, the spin phase the
gravitomagnetic sector, and the electromagnetic phase the
curvature-induced response of the background gauge field, obtained by
solving the detector-frame Maxwell equations in the long-wavelength
regime. Despite their different origins, all contributions are
controlled by the same {tidal scale,
$|\ddot h|\sim\Omega_{\rm gw}^2h_0$, corresponding to the local
curvature $R\sim|\ddot h|/c^2$,} so the interferometer probes
spacetime curvature rather than the metric perturbation itself.

For a square Mach--Zehnder interferometer, each channel is filtered by a
geometry-dependent kernel, and in the low-frequency regime
$\Omega_{\rm gw}\tau\ll1$ the responses obey simple scaling laws. The
dynamical phase provides the leading contribution among the intrinsic,
field-free channels, scales with the particle mass and interferometer
size, and reproduces the established tidal matter-wave response.
The spin phase encodes quantum-relativistic effects but is suppressed
by the ratio {$\lambdabar_C/L\ll1$}, rendering it unobservable via this
mechanism for any realistic apparatus; it is therefore a conceptual
consequence of the unified formalism rather than a candidate detection
channel. The electromagnetic phase is a different type of response: the GW first
induces space- and time-dependent perturbations of the background field,
and the resulting phase is set by the applied magnetic flux rather than
by the particle mass or spin. Within the initial-value solution and
geometry adopted here, this response depends on the orientation of the
background field relative to the GW propagation direction: the
perpendicular configuration carries an additional factor $c/v$ relative
to the parallel one, and the two configurations select different GW
polarizations. The phase is sensitive not only to the enclosed flux but
also to spatial gradients of the induced fields, in contrast to the
static Aharonov--Bohm effect.

Representative order-of-magnitude estimates illustrate the resulting
hierarchy: the spin channel is negligible, the parallel electromagnetic
response is comparable to the dynamical one for small flux areas, and,
for the representative parameters considered here, the perpendicular
electromagnetic response exceeds the dynamical channel by several orders
of magnitude, owing to the $c/v$ factor and the externally controlled
flux. These estimates follow from the low-frequency scaling laws under
the guided-arm approximation and do not represent the sensitivity of a
realizable detector; for charged particles, Lorentz-force dynamics,
electromagnetic decoherence, {spin-dependent forces from the guiding
mechanism,} and the boundary conditions of a confined
flux would all need to be addressed in a practical implementation.

Taken together, these results show that a single matter-wave
interferometer provides simultaneous access to gravitoelectric,
gravitomagnetic, and electromagnetically mediated aspects of spacetime
curvature. The present formulation places proper-time evolution, spin
transport, and electromagnetic holonomy within one covariant Dirac
framework, and provides a starting point for studying the interplay
between gravity, gauge fields, and quantum coherence in matter-wave
interferometry.

\begin{acknowledgments}
This work was supported by the Fundamental Fund of Thailand Science Research and Innovation (TSRI) through the National Astronomical Research Institute of Thailand (Public Organization) under grant FFB680072/0269, and by the National Science, Research and Innovation Fund (NSRF) via the Program Management Unit for Human Resources and Institutional Development, Research and Innovation under grant B39G680009.
\end{acknowledgments}

\section*{Data availability statement}
No new data were created or analysed in this study.

\appendix

\section{Maxwell equations in the detector frame: detailed derivation}
\label{app:maxwell_fnc_vacuum_gw}

Starting from \eqref{eq:maxwell_vacuum_main}, we expand the covariant
d'Alembertian in FNC to linear order in $h_{\mu\nu}$,
\begin{align}
\nabla^\rho \nabla_\rho A_\mu = &\;\square A_\mu
- h^{\alpha\beta} \partial_\alpha \partial_\beta A_\mu
- \mathcal{C}^\lambda \partial_\lambda A_\mu
\notag\\
&\; - \mathcal{D}_\mu{}^{\beta\lambda} \partial_\beta A_\lambda
- \tfrac{1}{2} \mathcal{P}_\mu{}^\lambda A_\lambda,
\label{eq:boxA_general_linear_app}
\end{align}
with $\square = -c^{-2}\partial_t^2 + \delta^{ij}\partial_i\partial_j$.
Here $\partial_0\equiv\partial/\partial x^0=(1/c)\partial_t$, a dot
denotes $d/dt$, and all indices on the perturbation and derivatives are
raised and lowered with the background Minkowski metric
$\eta_{\mu\nu}$; for purely spatial indices, this reduces to
$\delta_{ij}$. The coefficients are defined by
\begin{align}
\mathcal{C}^\lambda  &\equiv
\partial_\alpha h^{\alpha\lambda}
-\tfrac{1}{2}\partial^\lambda h,
\label{eq:C_operator_app}
\\[4pt]
\mathcal{D}_\mu{}^{\beta\lambda}
&\equiv
\partial_\mu h^{\beta\lambda}
+
\partial^\beta h_\mu{}^{\lambda}
-
\partial^\lambda h_\mu{}^{\beta},
\label{eq:D_operator_app}\\
\mathcal{P}_\mu{}^\lambda
&\equiv
\partial^\beta \partial_\mu h_\beta{}^{\lambda}
+
\square h_\mu{}^{\lambda}
-
\partial^\lambda \partial^\beta h_{\mu\beta},
\label{eq:P_operator_app}
\end{align}
where $h \equiv h^\alpha{}_\alpha$.

As stated in Sec.~\ref{sec:AB_GW}, we further work in the spatially uniform
regime, $\frac{L}{\lambdabar_{\rm gw}}\ll1$, so that spatial variations of the tidal fields across the apparatus are negligible. Accordingly, the tidal tensors are evaluated on the reference world line and treated as functions of time only,
with $\partial_k\mathcal E_{ij} = \partial_k\mathcal B_{ij}\simeq0.$

Using \eqref{eq:fnc_metric_tidal_main}, the derivatives of $h_{\mu\nu}$ are
\begin{subequations}
\begin{align}
\partial_0 h_{00}
&=
-\frac{1}{c}\,\dot{\mathcal{E}}_{ij}\,x^i x^j,
\\
\partial_m h_{00}
&=
-2\,\mathcal{E}_{mj}\,x^j,
\\
\partial_0 h_{0i}
&=
-\frac{2}{3c}\,\epsilon_{ik}{}^{l}\,\dot{\mathcal{B}}_{lj}\,x^j x^k,
\\
\partial_m h_{0i}
&=
-\frac{2}{3}
\left(
\epsilon_{im}{}^{l}\,\mathcal{B}_{lj}\,x^j
+
\epsilon_{ik}{}^{l}\,\mathcal{B}_{lm}\,x^k
\right),
\\
\partial_0 h_{ij}
&=
\frac{1}{3c}\,
\epsilon_{ik}{}^{m}\,\epsilon_{jl}{}^{n}\,
\dot{\mathcal{E}}_{mn}\,x^k x^l,
\\
\partial_p h_{ij}
&=
\frac{1}{3}
\left(
\epsilon_{ip}{}^{m}\,\epsilon_{jl}{}^{n}\,\mathcal{E}_{mn}\,x^l
+
\epsilon_{ik}{}^{m}\,\epsilon_{jp}{}^{n}\,\mathcal{E}_{mn}\,x^k
\right).
\end{align}
\end{subequations}

Substituting these expressions into the definitions of
$\mathcal C^\lambda$, $\mathcal D_\mu{}^{\beta\lambda}$, and
$\mathcal P_\mu{}^\lambda$ yields the explicit curvature-dependent
coefficients. {Two algebraic simplifications are used repeatedly
in this reduction. First, since $\mathcal E_{ij}$ and $\mathcal B_{ij}$
are symmetric in vacuum, any term of the form
$\epsilon_{ijk}\mathcal E_{jk}$ or $\epsilon_{ijk}\mathcal B_{jk}$
vanishes identically, as the antisymmetric $\epsilon$-tensor annihilates
a symmetric contraction. Second, the vacuum (Ricci-flat) condition
implies $\mathcal E_{ii}=\mathcal B_{ii}=0$, so any trace term
$\delta^{ij}\mathcal E_{ij}$ or $\delta^{ij}\mathcal B_{ij}$
arising from index contractions is discarded. Applying these
identities to $\partial_ih^{im}$ and $\partial_ih^{i0}$ above, together
with the $\epsilon\epsilon\to\delta\delta$ identity
$\epsilon_{iab}\epsilon_{icd}=\delta_{ac}\delta_{bd}-\delta_{ad}\delta_{bc}$
for the quadratic terms, one finds, for example,}
\begin{equation}
\mathcal C^0
=
-\frac{2}{3c}\dot{\mathcal E}_{ij}x^ix^j,
\qquad
\mathcal C^m
=
\frac{2}{3c}\epsilon_{mk}{}^l
\dot{\mathcal B}_{lj}x^jx^k
-\frac{1}{3}\mathcal E_{mj}x^j.
\label{eq:C_components_app}
\end{equation}

The correction to the principal second-derivative operator is
\begin{equation}
\begin{aligned}
-h^{\alpha\beta}\partial_\alpha\partial_\beta A_\mu
&=
\mathcal E_{ij}x^ix^j
\frac{1}{c^2}\partial_t^2A_\mu
\\
&\quad
-\frac{4}{3}
\epsilon_{ik}{}^l\mathcal B_{lj}x^jx^k
\frac{1}{c}\partial_t\partial_iA_\mu
\\
&\quad
-\frac{1}{3}
\epsilon_{ik}{}^m\epsilon_{jl}{}^n
\mathcal E_{mn}x^kx^l
\partial_i\partial_jA_\mu.
\end{aligned}
\label{eq:second_derivative_term_app}
\end{equation}

Combining these terms with the corresponding contributions from
$\mathcal D_\mu{}^{\beta\lambda}$ and
$\mathcal P_\mu{}^\lambda$ yields the coupled detector-frame equations for
the temporal and spatial components, $A_0$ and $A_i$.

\paragraph{Temporal component.}

Setting $\mu=0$ in Eq.~\eqref{eq:boxA_general_linear_app}, one has
\begin{align}
    \nabla^\rho\nabla_\rho A_0 = &\;\square A_0
- h^{\alpha\beta}\partial_\alpha\partial_\beta A_0
- \mathcal{C}^\lambda\partial_\lambda A_0
\notag\\
&\; - \mathcal{D}_0{}^{\beta\lambda}\partial_\beta A_\lambda
- \tfrac{1}{2}\mathcal{P}_0{}^\lambda A_\lambda .
\label{eq:boxA0_start_app}
\end{align}
The coefficients governing the first-derivative mixing are
\begin{subequations}
\label{eq:D0_all_app}
\begin{align}
\mathcal{D}_0{}^{00}
&=
-\frac{1}{c}\dot{\mathcal{E}}_{ij}x^i x^j,
\\
\mathcal{D}_0{}^{i0}
&=
2\mathcal{E}_{ij}x^j,
\\
\mathcal{D}_0{}^{0i}
&=
\frac{4}{3c}\epsilon_{ik}{}^{l}
\dot{\mathcal{B}}_{lj}x^j x^k
-
2\mathcal{E}_{ij}x^j,
\\
\mathcal{D}_0{}^{ij}
&=
\frac{1}{3c}
\epsilon_{ik}{}^{m}\epsilon_{jl}{}^{n}
\dot{\mathcal{E}}_{mn}x^k x^l
+
\frac{4}{3}\epsilon_{ij}{}^{l}\mathcal{B}_{lm}x^m
\nonumber\\
&\quad
-
\frac{2}{3}
\left(
\epsilon_{jk}{}^{l}\mathcal{B}_{li}
-
\epsilon_{ik}{}^{l}\mathcal{B}_{lj}
\right)x^k .
\end{align}
\end{subequations}
The algebraic coupling terms $\mathcal{P}_0{}^\lambda$ follow from second
derivatives of the metric perturbation. Using the identities above, together
with
\begin{equation}
\partial_m h_{0m}=0,\qquad
\partial_m\partial_m h_0{}^i=0,\qquad
{
\partial_m\partial_0 h_m{}^i
=
\frac{1}{3c}\dot{\mathcal{E}}_{ij}x^j,
}
\end{equation}
one obtains
\begin{align}
\mathcal{P}_0{}^0 &= -\frac{1}{c^2}\ddot{\mathcal{E}}_{ij}x^i x^j,\\
\mathcal{P}_0{}^i &=
\frac{4}{3c^2}
\epsilon_{ik}{}^{l}
\ddot{\mathcal{B}}_{lj}x^j x^k
-
\frac{5}{3c}
\dot{\mathcal{E}}_{ij}x^j .
\label{eq:P0_components_app}
\end{align}

Substituting these results into Eq.~\eqref{eq:boxA0_start_app}, together with
the vacuum Maxwell equation~\eqref{eq:maxwell_vacuum_main}, yields the explicit
wave equation for $A_0$:
\begin{widetext}
\begin{equation}
\begin{aligned}
-\mu_0 J_0 &= \square A_0 + \mathcal{E}_{ij} x^i x^j \,\frac{1}{c^2}\partial_t^2 A_0 - \frac{4}{3}\,\epsilon_{ik}{}^{l}\mathcal{B}_{lj} x^j x^k \,\frac{1}{c}\,\partial_t \partial_i A_0 \\
&\quad - \frac{1}{3}\epsilon_{ik}{}^{m}\epsilon_{jl}{}^{n}
\mathcal{E}_{mn} x^k x^l \partial_i \partial_j A_0 + \frac{5}{3c}\dot{\mathcal{E}}_{ij} x^i x^j \partial_0 A_0 - \left(
\frac{2}{3c}\epsilon_{mk}{}^{l}\dot{\mathcal{B}}_{lj} x^j x^k
+ \frac{5}{3}\mathcal{E}_{mj} x^j \right)\partial_m A_0 \\
&\quad
- \left( \frac{4}{3c}\,\epsilon_{ik}{}^{l}\dot{\mathcal{B}}_{lj} x^j x^k
- 2\,\mathcal{E}_{ij} x^j \right)\partial_0 A_i - \biggl[
\frac{1}{3c}\epsilon_{jk}{}^{m}\epsilon_{il}{}^{n}
\dot{\mathcal{E}}_{mn} x^k x^l + \frac{4}{3}\,\epsilon_{ji}{}^{l}\mathcal{B}_{lm} x^m\\
&\qquad
- \frac{2}{3}\left(\epsilon_{ik}{}^{l}\mathcal{B}_{lj}
- \epsilon_{jk}{}^{l}\mathcal{B}_{li}\right) x^k
\biggr] \partial_j A_i + \frac{1}{2c^2}\ddot{\mathcal{E}}_{ij} x^i x^j A_0
-\left(
\frac{2}{3c^2}\epsilon_{ik}{}^{l}
\ddot{\mathcal{B}}_{lj}x^j x^k - \frac{5}{6c}\dot{\mathcal{E}}_{ij}x^j\right)A_i .
\end{aligned}
\label{eq:maxwell_A0_tidal_app}
\end{equation}

\paragraph{Spatial components.}
Setting $\mu=k$ in Eq.~\eqref{eq:boxA_general_linear_app}, the wave equation
reads
\begin{align}
\nabla^\rho\nabla_\rho A_k
={}&
\square A_k
-h^{\alpha\beta}\partial_\alpha\partial_\beta A_k
-\mathcal C^\lambda\partial_\lambda A_k
\notag\\
&-\mathcal D_k{}^{\beta\lambda}\partial_\beta A_\lambda
-\frac12\mathcal P_k{}^\lambda A_\lambda .
\label{eq:boxAk_start_app}
\end{align}

The derivative-mixing coefficients are
\begin{subequations}
\label{eq:Dk_all_app}
\begin{align}
\mathcal D_k{}^{00}
&=
-2\mathcal E_{kj}x^j,
\\
\mathcal D_k{}^{0i}
&=
-\frac{1}{3c}
\epsilon_{ka}{}^{m}\epsilon_{ib}{}^{n}
\dot{\mathcal E}_{mn}x^a x^b
+
\frac{2}{3}
\bigl[
\epsilon_{ik}{}^{l}\mathcal B_{lp}
+
\epsilon_{ip}{}^{l}\mathcal B_{lk}
\notag\\
&\hspace{3.0cm}
-
\epsilon_{ki}{}^{l}\mathcal B_{lp}
-
\epsilon_{kp}{}^{l}\mathcal B_{li}
\bigr]x^p,
\\
\mathcal D_k{}^{i0}
&=
\frac{1}{3c}
\epsilon_{ka}{}^{m}\epsilon_{ib}{}^{n}
\dot{\mathcal E}_{mn}x^a x^b
+
\frac{2}{3}
\bigl[
\epsilon_{ik}{}^{l}\mathcal B_{lp}
+
\epsilon_{ip}{}^{l}\mathcal B_{lk}
\notag\\
&\hspace{3.0cm}
+
\epsilon_{ki}{}^{l}\mathcal B_{lp}
+
\epsilon_{kp}{}^{l}\mathcal B_{li}
\bigr]x^p,
\\
\mathcal D_k{}^{ij}
&=
\frac{1}{3}
\bigl[
\epsilon_{ik}{}^{m}\epsilon_{jl}{}^{n}
+
\epsilon_{il}{}^{m}\epsilon_{jk}{}^{n}
+
\epsilon_{ki}{}^{m}\epsilon_{jl}{}^{n}
\notag\\
&\hspace{1.5cm}
+
\epsilon_{kl}{}^{m}\epsilon_{ji}{}^{n}
-
\epsilon_{kj}{}^{m}\epsilon_{il}{}^{n}
-
\epsilon_{kl}{}^{m}\epsilon_{ij}{}^{n}
\bigr]
\mathcal E_{mn}x^l .
\end{align}
\end{subequations}

The algebraic coefficients are
\begin{equation}
{
\mathcal P_k{}^0
=
-\frac{5}{3c}\dot{\mathcal E}_{kj}x^j,
}
\end{equation}
and
\begin{equation}
{
\begin{aligned}
\mathcal P_k{}^i
={}&
-\frac{2}{3}\mathcal E_{ki}
-\frac{1}{3c^2}
\epsilon_{ka}{}^{m}\epsilon_{ib}{}^{n}
\ddot{\mathcal E}_{mn}x^a x^b
\\
&+
\frac{2}{c}
\left(
\epsilon_{ip}{}^{l}\dot{\mathcal B}_{lk}
-
\epsilon_{kp}{}^{l}\dot{\mathcal B}_{li}
\right)x^p .
\end{aligned}
}
\label{eq:Pk_spatial_app}
\end{equation}

Substitution into Eq.~\eqref{eq:boxAk_start_app} yields
\begin{equation}
\begin{aligned}
-\mu_0J_k &= \square A_k
+ \mathcal E_{ij}x^ix^j\frac{1}{c^2}\partial_t^2 A_k
- \frac{4}{3}\epsilon_{ip}{}^l\mathcal B_{lj}x^jx^p\frac{1}{c}\partial_t\partial_iA_k \\
&\quad
-\frac{1}{3}\epsilon_{im}{}^a\epsilon_{jn}{}^b \mathcal E_{ab}x^mx^n
\partial_i\partial_jA_k
+ \frac{2}{3c}\dot{\mathcal E}_{ij}x^ix^j\partial_0A_k \\
&\quad
+ \left(
\frac{1}{3}\mathcal E_{mj}x^j
-\frac{2}{3c}
\epsilon_{mp}{}^l\dot{\mathcal B}_{lj}x^jx^p
\right)\partial_mA_k \\
&\quad
+2\mathcal E_{kj}x^j\partial_0A_0
-\mathcal D_k{}^{i0}\partial_iA_0
-\mathcal D_k{}^{0i}\partial_0A_i
-\mathcal D_k{}^{ji}\partial_jA_i \\
&\quad
+\frac{5}{6c}\dot{\mathcal E}_{kj}x^jA_0 \\
&\quad
+ \left[\frac{1}{3}\mathcal E_{ki}
+ \frac{1}{6c^2}
\epsilon_{ka}{}^{m}\epsilon_{ib}{}^{n}
\ddot{\mathcal E}_{mn}x^a x^b
\right. \\
&\qquad
\left.
-\frac{1}{c}
\left(
\epsilon_{ip}{}^{l}\dot{\mathcal B}_{lk}
- \epsilon_{kp}{}^{l}\dot{\mathcal B}_{li} \right)x^p
\right]A_i .
\end{aligned}
\label{eq:maxwell_Ak_tidal_app}
\end{equation}
\end{widetext}

\paragraph{Lorenz gauge condition}
The linearized Lorenz-gauge condition,
$\partial_\mu A^\mu-h^{\mu\nu}\partial_\mu A_\nu
-\mathcal C^\nu A_\nu=0$, becomes
\begin{equation}
\begin{aligned}
-\partial_0 A_0 + \partial_i A_i
&+ \mathcal{E}_{ij}x^i x^j \partial_0 A_0
- \frac{2}{3}\epsilon_{ik}{}^{l}\mathcal{B}_{lj}x^j x^k
\partial_0 A_i
\\
&- \frac{2}{3}\epsilon_{ik}{}^{l}\mathcal{B}_{lj}x^j x^k
\partial_i A_0
- \frac{1}{3}\epsilon_{ik}{}^{m}\epsilon_{jl}{}^{n}
\mathcal{E}_{mn}x^k x^l \partial_i A_j
\\
&{
+ \frac{2}{3c}\dot{\mathcal{E}}_{ij}x^i x^j A_0
+ \left(
-\frac{2}{3c}\epsilon_{ik}{}^{l}
\dot{\mathcal{B}}_{lj}x^j x^k
+ \frac{1}{3}\mathcal{E}_{ij}x^j
\right)A_i
}
=0 .
\end{aligned}
\label{eq:lorenz_gauge_tidal_app}
\end{equation}

\section{Perturbative solution of the Maxwell--Lorenz system}
\label{app:solution_perp}

This appendix presents the explicit solution of the reduced Maxwell--Lorenz
system introduced in Sec.~\ref{sec:EM_response}, for both the
perpendicular ($\mathbf n=\hat{\mathbf x}$) and parallel
($\mathbf n=\hat{\mathbf z}$) configurations. The same conventions as in
Sec.~\ref{sec:EM_response} are adopted:
{the long-wavelength regime $L\ll\lambdabar_{\rm gw}$, in which all
time derivatives of the tidal tensors are dropped from the field
equations,} the background field $\mathbf B^{(0)}=B_0\hat{\mathbf z}$,
{and $\mathcal E_A=-\ddot h_A/(2c^2)$.} The reduced equations and
gauge condition are given in Eqs.~\eqref{eq:reduced_wave_main} and
\eqref{eq:lorenz_reduced_main_text}.

{
Throughout, coefficients are first obtained for arbitrary time dependence
of the tidal tensors via the double time
integral~\eqref{eq:double_integral_solution}; the closed-form expressions
below correspond to its short-time evaluation, in which
$\mathcal E_A$ is treated as constant over the integration interval.
Time derivatives of the tidal tensors generated by this procedure (e.g.,
$\ddot f\propto\dot{\mathcal E}_A$) are of the same dotted-suppressed
order as the terms already neglected in the field equations, so the
solution is consistent at the order retained.
}

\subsection{Perpendicular configuration}

{
For $\mathbf n=\hat{\mathbf x}$, the sources are
$S_0=-2B_0(\mathcal E_+y+\mathcal E_\times z)$,
the components of Eq.~\eqref{eq:Sk_perp_components}, and the gauge source
$G$ of Eq.~\eqref{eq:G_perp_main}.
}
The polynomial structure of the sources implies that $A_\mu^{(1)}$ is at
most quadratic in the spatial coordinates, and the ansatz is taken as
\begin{align}
A_0^{(1)} &= \alpha(t)y + \beta(t)z + p(t)xy + q(t)xz,
\nonumber\\
A_x^{(1)} &= r(t)y + s(t)z,
\nonumber\\
A_y^{(1)} &= m(t)x + f(t)(y^2 - x^2),
\nonumber\\
A_z^{(1)} &= n(t)x + g(t)(z^2 - x^2).
\end{align}
All monomials appearing here are harmonic, so the spatial Laplacian drops
out and substitution into Eq.~\eqref{eq:reduced_wave_main} with
$\square=-c^{-2}\partial_t^2+\nabla^2$ yields
{
\begin{align}
\ddot \alpha &= 2c^2 B_0 \mathcal E_+, &
\ddot \beta &= 2c^2 B_0 \mathcal E_\times,
\nonumber\\
\ddot r &= -\frac{7c^2B_0}{6}\mathcal E_+, &
\ddot s &= -\frac{7c^2B_0}{6}\mathcal E_\times,
\nonumber\\
\ddot m &= \frac{7c^2B_0}{6}\mathcal E_+, &
\ddot n &= \frac{7c^2B_0}{6}\mathcal E_\times,
\nonumber\\
\ddot p &= \ddot q = 0, &
\ddot f &= \ddot g = 0 .
\end{align}
}
The gauge condition~\eqref{eq:lorenz_reduced_main_text}, with
$\partial_0=(1/c)\partial_t$, gives upon matching monomials
{
\begin{equation}
f = \frac{\dot\alpha}{2c},
\qquad
g = \frac{\dot\beta}{2c},
\qquad
\dot p = \frac{cB_0}{6}\mathcal E_+,
\qquad
\dot q = \frac{cB_0}{6}\mathcal E_\times .
\end{equation}
}
These relations enforce consistency between the scalar and vector
potentials.

{
Imposing vanishing initial conditions and expressing the result in terms
of $\ddot h_A=-2c^2\mathcal E_A$, the solution reads
\begin{align}
A_0^{(1)} &=
-\frac{B_0 t^2}{2}\left(\ddot h_+ y+\ddot h_\times z\right)
-\frac{B_0 t}{12c}\,x\left(\ddot h_+ y+\ddot h_\times z\right),
\nonumber\\
A_x^{(1)} &=
\frac{7B_0 t^2}{24}\left(\ddot h_+ y+\ddot h_\times z\right),
\nonumber\\
A_y^{(1)} &=
-\frac{7B_0 t^2}{24}\,\ddot h_+\,x
-\frac{B_0 t}{2c}\,\ddot h_+\left(y^2-x^2\right),
\nonumber\\
A_z^{(1)} &=
-\frac{7B_0 t^2}{24}\,\ddot h_\times\,x
-\frac{B_0 t}{2c}\,\ddot h_\times\left(z^2-x^2\right).
\end{align}
Computing $E_i=-\partial_tA_i+c\,\partial_iA_0$ and
$B_k=\tfrac12\epsilon_{kij}(\partial_iA_j-\partial_jA_i)$ from these
potentials reproduces the fields quoted in
Sec.~\ref{sec:EM_response}; for example, the two contributions to
$\delta E_x$ combine as $-\tfrac{7}{12}-\tfrac{1}{12}=-\tfrac{2}{3}$.
}

\subsection{{Parallel configuration}}

{
For $\mathbf n=\hat{\mathbf z}$, all wave-equation sources vanish,
$S_0=S_k=0$, and the response is driven entirely by the gauge source
$G$ of Eq.~\eqref{eq:G_para_main}. The consistent ansatz is purely
scalar,
\begin{equation}
A_i^{(1)}=0,
\qquad
A_0^{(1)} = u(t)\,xy + v(t)\left(y^2-x^2\right),
\end{equation}
where both monomials are harmonic, so that
$\square A_0^{(1)}=0$ is satisfied identically with $\ddot u=\ddot v=0$.
The gauge condition fixes
\begin{equation}
\dot u = -\frac{cB_0}{3}\mathcal E_+,
\qquad
\dot v = -\frac{cB_0}{6}\mathcal E_\times ,
\end{equation}
and vanishing initial conditions give
\begin{equation}
A_0^{(1)}
=
-\frac{cB_0\,t}{6}
\left[
2\mathcal E_+\,xy
+
\mathcal E_\times\left(y^2-x^2\right)
\right],
\end{equation}
in agreement with Sec.~\ref{sec:EM_response}. Since $A_i^{(1)}=0$, the
response is purely electric, $\delta E_i=c\,\partial_iA_0^{(1)}$, and
$\delta\mathbf B=\mathbf 0$.
}

{
In both configurations, the solution contains homogeneous contributions
proportional to $t^2$ together with spatially varying terms proportional
to $xt$ and quadratic monomials. The latter arise from the interplay
between tidal driving and the Lorenz gauge constraint, and their
interpretation and range of validity are discussed at the end of
Sec.~\ref{sec:EM_response}.
}

\bibliographystyle{apsrev4-2}

\begin{thebibliography}{46}%
\makeatletter
\providecommand \@ifxundefined [1]{%
 \@ifx{#1\undefined}
}%
\providecommand \@ifnum [1]{%
 \ifnum #1\expandafter \@firstoftwo
 \else \expandafter \@secondoftwo
 \fi
}%
\providecommand \@ifx [1]{%
 \ifx #1\expandafter \@firstoftwo
 \else \expandafter \@secondoftwo
 \fi
}%
\providecommand \natexlab [1]{#1}%
\providecommand \enquote  [1]{``#1''}%
\providecommand \bibnamefont  [1]{#1}%
\providecommand \bibfnamefont [1]{#1}%
\providecommand \citenamefont [1]{#1}%
\providecommand \href@noop [0]{\@secondoftwo}%
\providecommand \href [0]{\begingroup \@sanitize@url \@href}%
\providecommand \@href[1]{\@@startlink{#1}\@@href}%
\providecommand \@@href[1]{\endgroup#1\@@endlink}%
\providecommand \@sanitize@url [0]{\catcode `\\12\catcode `\$12\catcode `\&12\catcode `\#12\catcode `\^12\catcode `\_12\catcode `\%12\relax}%
\providecommand \@@startlink[1]{}%
\providecommand \@@endlink[0]{}%
\providecommand \url  [0]{\begingroup\@sanitize@url \@url }%
\providecommand \@url [1]{\endgroup\@href {#1}{\urlprefix }}%
\providecommand \urlprefix  [0]{URL }%
\providecommand \Eprint [0]{\href }%
\providecommand \doibase [0]{https://doi.org/}%
\providecommand \selectlanguage [0]{\@gobble}%
\providecommand \bibinfo  [0]{\@secondoftwo}%
\providecommand \bibfield  [0]{\@secondoftwo}%
\providecommand \translation [1]{[#1]}%
\providecommand \BibitemOpen [0]{}%
\providecommand \bibitemStop [0]{}%
\providecommand \bibitemNoStop [0]{.\EOS\space}%
\providecommand \EOS [0]{\spacefactor3000\relax}%
\providecommand \BibitemShut  [1]{\csname bibitem#1\endcsname}%
\let\auto@bib@innerbib\@empty

\bibitem [{\citenamefont {Aharonov}\ and\ \citenamefont {Bohm}(1959)}]{Aharonov1959}%
  \BibitemOpen
  \bibfield  {author} {\bibinfo {author} {\bibfnamefont {Y.}~\bibnamefont {Aharonov}}\ and\ \bibinfo {author} {\bibfnamefont {D.}~\bibnamefont {Bohm}},\ }\href {https://doi.org/10.1103/PhysRev.115.485} {\bibfield  {journal} {\bibinfo  {journal} {Phys. Rev.}\ }\textbf {\bibinfo {volume} {115}},\ \bibinfo {pages} {485} (\bibinfo {year} {1959})}\BibitemShut {NoStop}%
\bibitem [{\citenamefont {Berry}(1984)}]{Berry1984}%
  \BibitemOpen
  \bibfield  {author} {\bibinfo {author} {\bibfnamefont {M.~V.}\ \bibnamefont {Berry}},\ }\href {https://doi.org/10.1098/rspa.1984.0023} {\bibfield  {journal} {\bibinfo  {journal} {Proc. R. Soc. A}\ }\textbf {\bibinfo {volume} {392}},\ \bibinfo {pages} {45} (\bibinfo {year} {1984})}\BibitemShut {NoStop}%
\bibitem [{\citenamefont {Aharonov}\ and\ \citenamefont {Anandan}(1987)}]{AharonovAnandan1987}%
  \BibitemOpen
  \bibfield  {author} {\bibinfo {author} {\bibfnamefont {Y.}~\bibnamefont {Aharonov}}\ and\ \bibinfo {author} {\bibfnamefont {J.}~\bibnamefont {Anandan}},\ }\href {https://doi.org/10.1103/PhysRevLett.58.1593} {\bibfield  {journal} {\bibinfo  {journal} {Phys. Rev. Lett.}\ }\textbf {\bibinfo {volume} {58}},\ \bibinfo {pages} {1593} (\bibinfo {year} {1987})}\BibitemShut {NoStop}%
\bibitem [{\citenamefont {Tonomura}\ \emph {et~al.}(1986)\citenamefont {Tonomura}, \citenamefont {Osakabe}, \citenamefont {Matsuda}, \citenamefont {Kawasaki}, \citenamefont {Endo}, \citenamefont {Yano},\ and\ \citenamefont {Yamada}}]{Tonomura1986}%
  \BibitemOpen
  \bibfield  {author} {\bibinfo {author} {\bibfnamefont {A.}~\bibnamefont {Tonomura}}, \bibinfo {author} {\bibfnamefont {N.}~\bibnamefont {Osakabe}}, \bibinfo {author} {\bibfnamefont {T.}~\bibnamefont {Matsuda}}, \bibinfo {author} {\bibfnamefont {T.}~\bibnamefont {Kawasaki}}, \bibinfo {author} {\bibfnamefont {J.}~\bibnamefont {Endo}}, \bibinfo {author} {\bibfnamefont {S.}~\bibnamefont {Yano}},\ and\ \bibinfo {author} {\bibfnamefont {H.}~\bibnamefont {Yamada}},\ }\href {https://doi.org/10.1103/PhysRevLett.56.792} {\bibfield  {journal} {\bibinfo  {journal} {Phys. Rev. Lett.}\ }\textbf {\bibinfo {volume} {56}},\ \bibinfo {pages} {792} (\bibinfo {year} {1986})}\BibitemShut {NoStop}%
\bibitem [{\citenamefont {Colella}\ \emph {et~al.}(1975)\citenamefont {Colella}, \citenamefont {Overhauser},\ and\ \citenamefont {Werner}}]{colella1975}%
  \BibitemOpen
  \bibfield  {author} {\bibinfo {author} {\bibfnamefont {R.}~\bibnamefont {Colella}}, \bibinfo {author} {\bibfnamefont {A.~W.}\ \bibnamefont {Overhauser}},\ and\ \bibinfo {author} {\bibfnamefont {S.~A.}\ \bibnamefont {Werner}},\ }\href {https://doi.org/10.1103/PhysRevLett.34.1472} {\bibfield  {journal} {\bibinfo  {journal} {Phys. Rev. Lett.}\ }\textbf {\bibinfo {volume} {34}},\ \bibinfo {pages} {1472} (\bibinfo {year} {1975})}\BibitemShut {NoStop}%
\bibitem [{\citenamefont {Stodolsky}(1979)}]{Stodolsky1979}%
  \BibitemOpen
  \bibfield  {author} {\bibinfo {author} {\bibfnamefont {L.}~\bibnamefont {Stodolsky}},\ }\href {https://doi.org/10.1007/BF00759302} {\bibfield  {journal} {\bibinfo  {journal} {Gen. Relativ. Gravit.}\ }\textbf {\bibinfo {volume} {11}},\ \bibinfo {pages} {391} (\bibinfo {year} {1979})}\BibitemShut {NoStop}%
\bibitem [{\citenamefont {Asenbaum}\ \emph {et~al.}(2017)\citenamefont {Asenbaum}, \citenamefont {Overstreet}, \citenamefont {Kovachy}, \citenamefont {Brown}, \citenamefont {Hogan},\ and\ \citenamefont {Kasevich}}]{Asenbaum2017}%
  \BibitemOpen
  \bibfield  {author} {\bibinfo {author} {\bibfnamefont {P.}~\bibnamefont {Asenbaum}}, \bibinfo {author} {\bibfnamefont {C.}~\bibnamefont {Overstreet}}, \bibinfo {author} {\bibfnamefont {T.}~\bibnamefont {Kovachy}}, \bibinfo {author} {\bibfnamefont {D.~D.}\ \bibnamefont {Brown}}, \bibinfo {author} {\bibfnamefont {J.~M.}\ \bibnamefont {Hogan}},\ and\ \bibinfo {author} {\bibfnamefont {M.~A.}\ \bibnamefont {Kasevich}},\ }\href {https://doi.org/10.1103/PhysRevLett.118.183602} {\bibfield  {journal} {\bibinfo  {journal} {Phys. Rev. Lett.}\ }\textbf {\bibinfo {volume} {118}},\ \bibinfo {pages} {183602} (\bibinfo {year} {2017})}\BibitemShut {NoStop}%
\bibitem [{\citenamefont {Asenbaum}\ \emph {et~al.}(2020)\citenamefont {Asenbaum}, \citenamefont {Overstreet}, \citenamefont {Kim}, \citenamefont {Curti},\ and\ \citenamefont {Kasevich}}]{Asenbaum2020}%
  \BibitemOpen
  \bibfield  {author} {\bibinfo {author} {\bibfnamefont {P.}~\bibnamefont {Asenbaum}}, \bibinfo {author} {\bibfnamefont {C.}~\bibnamefont {Overstreet}}, \bibinfo {author} {\bibfnamefont {M.}~\bibnamefont {Kim}}, \bibinfo {author} {\bibfnamefont {J.}~\bibnamefont {Curti}},\ and\ \bibinfo {author} {\bibfnamefont {M.~A.}\ \bibnamefont {Kasevich}},\ }\href {https://doi.org/10.1103/PhysRevLett.125.191101} {\bibfield  {journal} {\bibinfo  {journal} {Phys. Rev. Lett.}\ }\textbf {\bibinfo {volume} {125}},\ \bibinfo {pages} {191101} (\bibinfo {year} {2020})}\BibitemShut {NoStop}%
\bibitem [{\citenamefont {Dimopoulos}\ \emph {et~al.}(2008)\citenamefont {Dimopoulos}, \citenamefont {Graham}, \citenamefont {Hogan}, \citenamefont {Kasevich},\ and\ \citenamefont {Rajendran}}]{Dimopoulos2008}%
  \BibitemOpen
  \bibfield  {author} {\bibinfo {author} {\bibfnamefont {S.}~\bibnamefont {Dimopoulos}}, \bibinfo {author} {\bibfnamefont {P.~W.}\ \bibnamefont {Graham}}, \bibinfo {author} {\bibfnamefont {J.~M.}\ \bibnamefont {Hogan}}, \bibinfo {author} {\bibfnamefont {M.~A.}\ \bibnamefont {Kasevich}},\ and\ \bibinfo {author} {\bibfnamefont {S.}~\bibnamefont {Rajendran}},\ }\href {https://doi.org/10.1103/PhysRevD.78.122002} {\bibfield  {journal} {\bibinfo  {journal} {Phys. Rev. D}\ }\textbf {\bibinfo {volume} {78}},\ \bibinfo {pages} {122002} (\bibinfo {year} {2008})}\BibitemShut {NoStop}%
\bibitem [{\citenamefont {Graham}\ \emph {et~al.}(2013)\citenamefont {Graham}, \citenamefont {Hogan}, \citenamefont {Kasevich},\ and\ \citenamefont {Rajendran}}]{Graham2013}%
  \BibitemOpen
  \bibfield  {author} {\bibinfo {author} {\bibfnamefont {P.~W.}\ \bibnamefont {Graham}}, \bibinfo {author} {\bibfnamefont {J.~M.}\ \bibnamefont {Hogan}}, \bibinfo {author} {\bibfnamefont {M.~A.}\ \bibnamefont {Kasevich}},\ and\ \bibinfo {author} {\bibfnamefont {S.}~\bibnamefont {Rajendran}},\ }\href {https://doi.org/10.1103/PhysRevLett.110.171102} {\bibfield  {journal} {\bibinfo  {journal} {Phys. Rev. Lett.}\ }\textbf {\bibinfo {volume} {110}},\ \bibinfo {pages} {171102} (\bibinfo {year} {2013})}\BibitemShut {NoStop}%
\bibitem [{\citenamefont {Canuel}\ \emph {et~al.}(2018)\citenamefont {Canuel}, \citenamefont {Bertoldi}, \citenamefont {Amand}, \citenamefont {Pozzo~di Borgo}, \citenamefont {Chantrait}, \citenamefont {Danquigny}, \citenamefont {Dovale~{\'A}lvarez}, \citenamefont {Fang}, \citenamefont {Freise}, \citenamefont {Geiger}, \citenamefont {Gillot}, \citenamefont {Henry}, \citenamefont {Hinderer}, \citenamefont {Holleville}, \citenamefont {Junca}, \citenamefont {Lef{\`e}vre}, \citenamefont {Merzougui}, \citenamefont {Mielec}, \citenamefont {Monfret}, \citenamefont {Pelisson}, \citenamefont {Prevedelli}, \citenamefont {Reynaud}, \citenamefont {Riou}, \citenamefont {Rogister}, \citenamefont {Rosat}, \citenamefont {Cormier}, \citenamefont {Landragin}, \citenamefont {Chaibi}, \citenamefont {Gaffet},\ and\ \citenamefont {Bouyer}}]{Canuel2018}%
  \BibitemOpen
  \bibfield  {author} {\bibinfo {author} {\bibfnamefont {B.}~\bibnamefont {Canuel}}, \bibinfo {author} {\bibfnamefont {A.}~\bibnamefont {Bertoldi}}, \bibinfo {author} {\bibfnamefont {L.}~\bibnamefont {Amand}}, \bibinfo {author} {\bibfnamefont {E.}~\bibnamefont {Pozzo~di Borgo}}, \bibinfo {author} {\bibfnamefont {T.}~\bibnamefont {Chantrait}}, \bibinfo {author} {\bibfnamefont {C.}~\bibnamefont {Danquigny}}, \bibinfo {author} {\bibfnamefont {M.}~\bibnamefont {Dovale~{\'A}lvarez}}, \bibinfo {author} {\bibfnamefont {B.}~\bibnamefont {Fang}}, \bibinfo {author} {\bibfnamefont {A.}~\bibnamefont {Freise}}, \bibinfo {author} {\bibfnamefont {R.}~\bibnamefont {Geiger}}, \bibinfo {author} {\bibfnamefont {J.}~\bibnamefont {Gillot}}, \bibinfo {author} {\bibfnamefont {S.}~\bibnamefont {Henry}}, \bibinfo {author} {\bibfnamefont {J.}~\bibnamefont {Hinderer}}, \bibinfo {author} {\bibfnamefont {D.}~\bibnamefont {Holleville}}, \bibinfo {author} {\bibfnamefont {J.}~\bibnamefont {Junca}}, \bibinfo {author} {\bibfnamefont
  {G.}~\bibnamefont {Lef{\`e}vre}}, \bibinfo {author} {\bibfnamefont {M.}~\bibnamefont {Merzougui}}, \bibinfo {author} {\bibfnamefont {N.}~\bibnamefont {Mielec}}, \bibinfo {author} {\bibfnamefont {T.}~\bibnamefont {Monfret}}, \bibinfo {author} {\bibfnamefont {S.}~\bibnamefont {Pelisson}}, \bibinfo {author} {\bibfnamefont {M.}~\bibnamefont {Prevedelli}}, \bibinfo {author} {\bibfnamefont {S.}~\bibnamefont {Reynaud}}, \bibinfo {author} {\bibfnamefont {I.}~\bibnamefont {Riou}}, \bibinfo {author} {\bibfnamefont {Y.}~\bibnamefont {Rogister}}, \bibinfo {author} {\bibfnamefont {S.}~\bibnamefont {Rosat}}, \bibinfo {author} {\bibfnamefont {E.}~\bibnamefont {Cormier}}, \bibinfo {author} {\bibfnamefont {A.}~\bibnamefont {Landragin}}, \bibinfo {author} {\bibfnamefont {W.}~\bibnamefont {Chaibi}}, \bibinfo {author} {\bibfnamefont {S.}~\bibnamefont {Gaffet}},\ and\ \bibinfo {author} {\bibfnamefont {P.}~\bibnamefont {Bouyer}},\ }\href {https://doi.org/10.1038/s41598-018-32165-z} {\bibfield  {journal} {\bibinfo  {journal}
  {Sci. Rep.}\ }\textbf {\bibinfo {volume} {8}},\ \bibinfo {pages} {14064} (\bibinfo {year} {2018})}\BibitemShut {NoStop}%
\bibitem [{\citenamefont {Graham}\ \emph {et~al.}(2017)\citenamefont {Graham}, \citenamefont {Hogan}, \citenamefont {Kasevich}, \citenamefont {Rajendran},\ and\ \citenamefont {Romani}}]{MAGIS2021}%
  \BibitemOpen
  \bibfield  {author} {\bibinfo {author} {\bibfnamefont {P.~W.}\ \bibnamefont {Graham}}, \bibinfo {author} {\bibfnamefont {J.~M.}\ \bibnamefont {Hogan}}, \bibinfo {author} {\bibfnamefont {M.~A.}\ \bibnamefont {Kasevich}}, \bibinfo {author} {\bibfnamefont {S.}~\bibnamefont {Rajendran}},\ and\ \bibinfo {author} {\bibfnamefont {R.~W.}\ \bibnamefont {Romani}},\ }\href@noop {} {\bibinfo {title} {Mid-band gravitational wave detection with precision atomic sensors}} (\bibinfo {year} {2017}),\ \Eprint {https://arxiv.org/abs/1711.02225} {arXiv:1711.02225 [physics.atom-ph]} \BibitemShut {NoStop}%
\bibitem [{\citenamefont {Anandan}(1977)}]{Anandan1977}%
  \BibitemOpen
  \bibfield  {author} {\bibinfo {author} {\bibfnamefont {J.}~\bibnamefont {Anandan}},\ }\href {https://doi.org/10.1103/PhysRevD.15.1448} {\bibfield  {journal} {\bibinfo  {journal} {Phys. Rev. D}\ }\textbf {\bibinfo {volume} {15}},\ \bibinfo {pages} {1448} (\bibinfo {year} {1977})}\BibitemShut {NoStop}%
\bibitem [{\citenamefont {Anandan}(1981)}]{Anandan1981}%
  \BibitemOpen
  \bibfield  {author} {\bibinfo {author} {\bibfnamefont {J.}~\bibnamefont {Anandan}},\ }\href {https://doi.org/10.1103/PhysRevD.24.338} {\bibfield  {journal} {\bibinfo  {journal} {Phys. Rev. D}\ }\textbf {\bibinfo {volume} {24}},\ \bibinfo {pages} {338} (\bibinfo {year} {1981})}\BibitemShut {NoStop}%
\bibitem [{\citenamefont {Mashhoon}(1988)}]{Mashhoon1988}%
  \BibitemOpen
  \bibfield  {author} {\bibinfo {author} {\bibfnamefont {B.}~\bibnamefont {Mashhoon}},\ }\href {https://doi.org/10.1103/PhysRevLett.61.2639} {\bibfield  {journal} {\bibinfo  {journal} {Phys. Rev. Lett.}\ }\textbf {\bibinfo {volume} {61}},\ \bibinfo {pages} {2639} (\bibinfo {year} {1988})}\BibitemShut {NoStop}%
\bibitem [{\citenamefont {Papini}(2008)}]{Papini2008}%
  \BibitemOpen
  \bibfield  {author} {\bibinfo {author} {\bibfnamefont {G.}~\bibnamefont {Papini}},\ }\href {https://doi.org/10.1007/s10714-007-0595-z} {\bibfield  {journal} {\bibinfo  {journal} {Gen. Relativ. Gravit.}\ }\textbf {\bibinfo {volume} {40}},\ \bibinfo {pages} {1117} (\bibinfo {year} {2008})}\BibitemShut {NoStop}%
\bibitem [{\citenamefont {Bord{\'e}}\ \emph {et~al.}(2001)\citenamefont {Bord{\'e}}, \citenamefont {Houard},\ and\ \citenamefont {Karasiewicz}}]{Borde2000}%
  \BibitemOpen
  \bibfield  {author} {\bibinfo {author} {\bibfnamefont {C.~J.}\ \bibnamefont {Bord{\'e}}}, \bibinfo {author} {\bibfnamefont {J.-C.}\ \bibnamefont {Houard}},\ and\ \bibinfo {author} {\bibfnamefont {A.}~\bibnamefont {Karasiewicz}},\ }in\ \href {https://doi.org/10.1007/3-540-40988-2_21} {\emph {\bibinfo {booktitle} {Gyros, Clocks, Interferometers{\ldots}: Testing Relativistic Gravity in Space}}},\ \bibinfo {editor} {edited by\ \bibinfo {editor} {\bibfnamefont {C.}~\bibnamefont {L{\"a}mmerzahl}}, \bibinfo {editor} {\bibfnamefont {C.~W.~F.}\ \bibnamefont {Everitt}},\ and\ \bibinfo {editor} {\bibfnamefont {F.~W.}\ \bibnamefont {Hehl}}}\ (\bibinfo  {publisher} {Springer},\ \bibinfo {address} {Berlin, Heidelberg},\ \bibinfo {year} {2001})\ pp.\ \bibinfo {pages} {403--438},\ \Eprint {https://arxiv.org/abs/gr-qc/0008033} {arXiv:gr-qc/0008033} \BibitemShut {NoStop}%
\bibitem [{\citenamefont {Javadinezhad}\ and\ \citenamefont {Seraj}(2026)}]{Javadinezhad2026}%
  \BibitemOpen
  \bibfield  {author} {\bibinfo {author} {\bibfnamefont {R.}~\bibnamefont {Javadinezhad}}\ and\ \bibinfo {author} {\bibfnamefont {A.}~\bibnamefont {Seraj}},\ }\href {https://doi.org/10.1007/JHEP05(2026)046} {\bibfield  {journal} {\bibinfo  {journal} {JHEP}\ }\textbf {\bibinfo {volume} {05}},\ \bibinfo {pages} {046}},\ \Eprint {https://arxiv.org/abs/2512.24429} {arXiv:2512.24429 [gr-qc]} \BibitemShut {NoStop}%
\bibitem [{\citenamefont {Audretsch}(1981)}]{Audretsch1981}%
  \BibitemOpen
  \bibfield  {author} {\bibinfo {author} {\bibfnamefont {J.}~\bibnamefont {Audretsch}},\ }\href {https://doi.org/10.1088/0305-4470/14/2/017} {\bibfield  {journal} {\bibinfo  {journal} {J. Phys. A: Math. Gen.}\ }\textbf {\bibinfo {volume} {14}},\ \bibinfo {pages} {411} (\bibinfo {year} {1981})}\BibitemShut {NoStop}%
\bibitem [{\citenamefont {Marklund}\ \emph {et~al.}(2000)\citenamefont {Marklund}, \citenamefont {Dunsby},\ and\ \citenamefont {Brodin}}]{Marklund2000}%
  \BibitemOpen
  \bibfield  {author} {\bibinfo {author} {\bibfnamefont {M.}~\bibnamefont {Marklund}}, \bibinfo {author} {\bibfnamefont {P.~K.~S.}\ \bibnamefont {Dunsby}},\ and\ \bibinfo {author} {\bibfnamefont {G.}~\bibnamefont {Brodin}},\ }\href {https://doi.org/10.1103/PhysRevD.62.101501} {\bibfield  {journal} {\bibinfo  {journal} {Phys. Rev. D}\ }\textbf {\bibinfo {volume} {62}},\ \bibinfo {pages} {101501} (\bibinfo {year} {2000})}\BibitemShut {NoStop}%
\bibitem [{\citenamefont {Brodin}\ \emph {et~al.}(2000)\citenamefont {Brodin}, \citenamefont {Marklund},\ and\ \citenamefont {Dunsby}}]{Brodin2001}%
  \BibitemOpen
  \bibfield  {author} {\bibinfo {author} {\bibfnamefont {G.}~\bibnamefont {Brodin}}, \bibinfo {author} {\bibfnamefont {M.}~\bibnamefont {Marklund}},\ and\ \bibinfo {author} {\bibfnamefont {P.~K.~S.}\ \bibnamefont {Dunsby}},\ }\href {https://doi.org/10.1103/PhysRevD.62.104008} {\bibfield  {journal} {\bibinfo  {journal} {Phys. Rev. D}\ }\textbf {\bibinfo {volume} {62}},\ \bibinfo {pages} {104008} (\bibinfo {year} {2000})}\BibitemShut {NoStop}%
\bibitem [{\citenamefont {Servin}\ and\ \citenamefont {Brodin}(2003)}]{Servin2003}%
  \BibitemOpen
  \bibfield  {author} {\bibinfo {author} {\bibfnamefont {M.}~\bibnamefont {Servin}}\ and\ \bibinfo {author} {\bibfnamefont {G.}~\bibnamefont {Brodin}},\ }\href {https://doi.org/10.1103/PhysRevD.68.044017} {\bibfield  {journal} {\bibinfo  {journal} {Phys. Rev. D}\ }\textbf {\bibinfo {volume} {68}},\ \bibinfo {pages} {044017} (\bibinfo {year} {2003})}\BibitemShut {NoStop}%
\bibitem [{\citenamefont {Domcke}\ \emph {et~al.}(2022)\citenamefont {Domcke}, \citenamefont {Garcia-Cely},\ and\ \citenamefont {Rodd}}]{Domcke2022}%
  \BibitemOpen
  \bibfield  {author} {\bibinfo {author} {\bibfnamefont {V.}~\bibnamefont {Domcke}}, \bibinfo {author} {\bibfnamefont {C.}~\bibnamefont {Garcia-Cely}},\ and\ \bibinfo {author} {\bibfnamefont {N.~L.}\ \bibnamefont {Rodd}},\ }\href {https://doi.org/10.1103/PhysRevLett.129.041101} {\bibfield  {journal} {\bibinfo  {journal} {Phys. Rev. Lett.}\ }\textbf {\bibinfo {volume} {129}},\ \bibinfo {pages} {041101} (\bibinfo {year} {2022})}\BibitemShut {NoStop}%
\bibitem [{\citenamefont {Ruggiero}(2025)}]{Ruggiero2025}%
  \BibitemOpen
  \bibfield  {author} {\bibinfo {author} {\bibfnamefont {M.~L.}\ \bibnamefont {Ruggiero}},\ }\href {https://doi.org/10.1140/epjc/s10052-025-13962-z} {\bibfield  {journal} {\bibinfo  {journal} {Eur. Phys. J. C}\ }\textbf {\bibinfo {volume} {85}},\ \bibinfo {pages} {240} (\bibinfo {year} {2025})}\BibitemShut {NoStop}%
\bibitem [{\citenamefont {Badurina}\ \emph {et~al.}(2025)\citenamefont {Badurina}, \citenamefont {Du}, \citenamefont {Lee}, \citenamefont {Wang},\ and\ \citenamefont {Zurek}}]{Badurina2025}%
  \BibitemOpen
  \bibfield  {author} {\bibinfo {author} {\bibfnamefont {L.}~\bibnamefont {Badurina}}, \bibinfo {author} {\bibfnamefont {Y.}~\bibnamefont {Du}}, \bibinfo {author} {\bibfnamefont {V.~S.~H.}\ \bibnamefont {Lee}}, \bibinfo {author} {\bibfnamefont {Y.}~\bibnamefont {Wang}},\ and\ \bibinfo {author} {\bibfnamefont {K.~M.}\ \bibnamefont {Zurek}},\ }\href {https://doi.org/10.1103/PhysRevD.111.042002} {\bibfield  {journal} {\bibinfo  {journal} {Phys. Rev. D}\ }\textbf {\bibinfo {volume} {111}},\ \bibinfo {pages} {042002} (\bibinfo {year} {2025})}\BibitemShut {NoStop}%
\bibitem [{\citenamefont {Schaffrath}\ \emph {et~al.}(2025)\citenamefont {Schaffrath}, \citenamefont {St{\"o}rk}, \citenamefont {Pumpo},\ and\ \citenamefont {Giese}}]{Schaffrath2025}%
  \BibitemOpen
  \bibfield  {author} {\bibinfo {author} {\bibfnamefont {S.}~\bibnamefont {Schaffrath}}, \bibinfo {author} {\bibfnamefont {D.}~\bibnamefont {St{\"o}rk}}, \bibinfo {author} {\bibfnamefont {F.~D.}\ \bibnamefont {Pumpo}},\ and\ \bibinfo {author} {\bibfnamefont {E.}~\bibnamefont {Giese}},\ }\href {https://doi.org/10.1116/5.0304468} {\bibfield  {journal} {\bibinfo  {journal} {AVS Quantum Sci.}\ }\textbf {\bibinfo {volume} {7}},\ \bibinfo {pages} {044402} (\bibinfo {year} {2025})}\BibitemShut {NoStop}%
\bibitem [{\citenamefont {Gaudio}(2026)}]{Gaudio2026}%
  \BibitemOpen
  \bibfield  {author} {\bibinfo {author} {\bibfnamefont {S.}~\bibnamefont {Gaudio}},\ }\href@noop {} {\bibinfo {title} {Fundamental limits of quantum sensors for gravitational wave detection}} (\bibinfo {year} {2026}),\ \Eprint {https://arxiv.org/abs/2603.06772} {arXiv:2603.06772 [gr-qc]} \BibitemShut {NoStop}%
\bibitem [{\citenamefont {Manasse}\ and\ \citenamefont {Misner}(1963)}]{ManasseMisner1963}%
  \BibitemOpen
  \bibfield  {author} {\bibinfo {author} {\bibfnamefont {F.~K.}\ \bibnamefont {Manasse}}\ and\ \bibinfo {author} {\bibfnamefont {C.~W.}\ \bibnamefont {Misner}},\ }\href {https://doi.org/10.1063/1.1724316} {\bibfield  {journal} {\bibinfo  {journal} {J. Math. Phys.}\ }\textbf {\bibinfo {volume} {4}},\ \bibinfo {pages} {735} (\bibinfo {year} {1963})}\BibitemShut {NoStop}%
\bibitem [{\citenamefont {Misner}\ \emph {et~al.}(1973)\citenamefont {Misner}, \citenamefont {Thorne},\ and\ \citenamefont {Wheeler}}]{MTW}%
  \BibitemOpen
  \bibfield  {author} {\bibinfo {author} {\bibfnamefont {C.~W.}\ \bibnamefont {Misner}}, \bibinfo {author} {\bibfnamefont {K.~S.}\ \bibnamefont {Thorne}},\ and\ \bibinfo {author} {\bibfnamefont {J.~A.}\ \bibnamefont {Wheeler}},\ }\href@noop {} {\emph {\bibinfo {title} {Gravitation}}}\ (\bibinfo  {publisher} {W. H. Freeman and Company},\ \bibinfo {address} {San Francisco},\ \bibinfo {year} {1973})\BibitemShut {NoStop}%
\bibitem [{\citenamefont {Poisson}\ \emph {et~al.}(2011)\citenamefont {Poisson}, \citenamefont {Pound},\ and\ \citenamefont {Vega}}]{PoissonPoundVega2011}%
  \BibitemOpen
  \bibfield  {author} {\bibinfo {author} {\bibfnamefont {E.}~\bibnamefont {Poisson}}, \bibinfo {author} {\bibfnamefont {A.}~\bibnamefont {Pound}},\ and\ \bibinfo {author} {\bibfnamefont {I.}~\bibnamefont {Vega}},\ }\href {https://doi.org/10.12942/lrr-2011-7} {\bibfield  {journal} {\bibinfo  {journal} {Living Rev. Relativ.}\ }\textbf {\bibinfo {volume} {14}},\ \bibinfo {pages} {7} (\bibinfo {year} {2011})}\BibitemShut {NoStop}%
\bibitem [{\citenamefont {Collas}\ and\ \citenamefont {Klein}(2019)}]{collas2019}%
  \BibitemOpen
  \bibfield  {author} {\bibinfo {author} {\bibfnamefont {P.}~\bibnamefont {Collas}}\ and\ \bibinfo {author} {\bibfnamefont {D.}~\bibnamefont {Klein}},\ }\href {https://doi.org/10.1007/978-3-030-14825-6} {\emph {\bibinfo {title} {The Dirac Equation in Curved Spacetime: A Guide for Calculations}}},\ SpringerBriefs in Physics\ (\bibinfo  {publisher} {Springer International Publishing},\ \bibinfo {address} {Cham},\ \bibinfo {year} {2019})\BibitemShut {NoStop}%
\bibitem [{\citenamefont {Parker}\ and\ \citenamefont {Toms}(2009)}]{ParkerToms2009}%
  \BibitemOpen
  \bibfield  {author} {\bibinfo {author} {\bibfnamefont {L.}~\bibnamefont {Parker}}\ and\ \bibinfo {author} {\bibfnamefont {D.~J.}\ \bibnamefont {Toms}},\ }\href@noop {} {\emph {\bibinfo {title} {Quantum Field Theory in Curved Spacetime: Quantized Fields and Gravity}}}\ (\bibinfo  {publisher} {Cambridge University Press},\ \bibinfo {address} {Cambridge},\ \bibinfo {year} {2009})\BibitemShut {NoStop}%
\bibitem [{\citenamefont {Oancea}\ and\ \citenamefont {Kumar}(2023)}]{Oancea2023}%
  \BibitemOpen
  \bibfield  {author} {\bibinfo {author} {\bibfnamefont {M.~A.}\ \bibnamefont {Oancea}}\ and\ \bibinfo {author} {\bibfnamefont {A.}~\bibnamefont {Kumar}},\ }\href {https://doi.org/10.1103/PhysRevD.107.044029} {\bibfield  {journal} {\bibinfo  {journal} {Phys. Rev. D}\ }\textbf {\bibinfo {volume} {107}},\ \bibinfo {pages} {044029} (\bibinfo {year} {2023})},\ \Eprint {https://arxiv.org/abs/2212.04414} {arXiv:2212.04414 [gr-qc]} \BibitemShut {NoStop}%
\bibitem [{\citenamefont {R{\"u}diger}(1981)}]{Rudiger1981}%
  \BibitemOpen
  \bibfield  {author} {\bibinfo {author} {\bibfnamefont {R.}~\bibnamefont {R{\"u}diger}},\ }\href {https://doi.org/10.1098/rspa.1981.0132} {\bibfield  {journal} {\bibinfo  {journal} {Proc. R. Soc. Lond. A}\ }\textbf {\bibinfo {volume} {377}},\ \bibinfo {pages} {417} (\bibinfo {year} {1981})}\BibitemShut {NoStop}%
\bibitem [{\citenamefont {Hammad}\ \emph {et~al.}(2024)\citenamefont {Hammad}, \citenamefont {Simard}, \citenamefont {Saadati},\ and\ \citenamefont {Landry}}]{Hammad2024}%
  \BibitemOpen
  \bibfield  {author} {\bibinfo {author} {\bibfnamefont {F.}~\bibnamefont {Hammad}}, \bibinfo {author} {\bibfnamefont {M.}~\bibnamefont {Simard}}, \bibinfo {author} {\bibfnamefont {R.}~\bibnamefont {Saadati}},\ and\ \bibinfo {author} {\bibfnamefont {A.}~\bibnamefont {Landry}},\ }\href {https://doi.org/10.1103/PhysRevD.110.065005} {\bibfield  {journal} {\bibinfo  {journal} {Phys. Rev. D}\ }\textbf {\bibinfo {volume} {110}},\ \bibinfo {pages} {065005} (\bibinfo {year} {2024})}\BibitemShut {NoStop}%
\bibitem [{\citenamefont {Ni}\ and\ \citenamefont {Zimmermann}(1978)}]{NiZimmermann1978}%
  \BibitemOpen
  \bibfield  {author} {\bibinfo {author} {\bibfnamefont {W.-T.}\ \bibnamefont {Ni}}\ and\ \bibinfo {author} {\bibfnamefont {M.}~\bibnamefont {Zimmermann}},\ }\href {https://doi.org/10.1103/PhysRevD.17.1473} {\bibfield  {journal} {\bibinfo  {journal} {Phys. Rev. D}\ }\textbf {\bibinfo {volume} {17}},\ \bibinfo {pages} {1473} (\bibinfo {year} {1978})}\BibitemShut {NoStop}%
\bibitem [{\citenamefont {Marzlin}(1994)}]{Marzlin1994}%
  \BibitemOpen
  \bibfield  {author} {\bibinfo {author} {\bibfnamefont {K.-P.}\ \bibnamefont {Marzlin}},\ }\href {https://doi.org/10.1103/PhysRevD.50.888} {\bibfield  {journal} {\bibinfo  {journal} {Phys. Rev. D}\ }\textbf {\bibinfo {volume} {50}},\ \bibinfo {pages} {888} (\bibinfo {year} {1994})}\BibitemShut {NoStop}%
\bibitem [{\citenamefont {Sagnac}(1913)}]{Sagnac1913}%
  \BibitemOpen
  \bibfield  {author} {\bibinfo {author} {\bibfnamefont {G.}~\bibnamefont {Sagnac}},\ }\href@noop {} {\bibfield  {journal} {\bibinfo  {journal} {Comptes Rendus de l'Acad{\'e}mie des Sciences}\ }\textbf {\bibinfo {volume} {157}},\ \bibinfo {pages} {708} (\bibinfo {year} {1913})}\BibitemShut {NoStop}%
\bibitem [{\citenamefont {Werner}\ \emph {et~al.}(1979)\citenamefont {Werner}, \citenamefont {Staudenmann},\ and\ \citenamefont {Colella}}]{Werner1979}%
  \BibitemOpen
  \bibfield  {author} {\bibinfo {author} {\bibfnamefont {S.~A.}\ \bibnamefont {Werner}}, \bibinfo {author} {\bibfnamefont {J.-L.}\ \bibnamefont {Staudenmann}},\ and\ \bibinfo {author} {\bibfnamefont {R.}~\bibnamefont {Colella}},\ }\href {https://doi.org/10.1103/PhysRevLett.42.1103} {\bibfield  {journal} {\bibinfo  {journal} {Phys. Rev. Lett.}\ }\textbf {\bibinfo {volume} {42}},\ \bibinfo {pages} {1103} (\bibinfo {year} {1979})}\BibitemShut {NoStop}%
\bibitem [{\citenamefont {Varj{\'u}}\ and\ \citenamefont {Ryder}(2000)}]{varju2000}%
  \BibitemOpen
  \bibfield  {author} {\bibinfo {author} {\bibfnamefont {K.}~\bibnamefont {Varj{\'u}}}\ and\ \bibinfo {author} {\bibfnamefont {L.~H.}\ \bibnamefont {Ryder}},\ }\href {https://doi.org/10.1103/PhysRevD.62.024016} {\bibfield  {journal} {\bibinfo  {journal} {Phys. Rev. D}\ }\textbf {\bibinfo {volume} {62}},\ \bibinfo {pages} {024016} (\bibinfo {year} {2000})}\BibitemShut {NoStop}%
\bibitem [{\citenamefont {Hehl}\ and\ \citenamefont {Ni}(1990)}]{HehlNi1990}%
  \BibitemOpen
  \bibfield  {author} {\bibinfo {author} {\bibfnamefont {F.~W.}\ \bibnamefont {Hehl}}\ and\ \bibinfo {author} {\bibfnamefont {W.-T.}\ \bibnamefont {Ni}},\ }\href {https://doi.org/10.1103/PhysRevD.42.2045} {\bibfield  {journal} {\bibinfo  {journal} {Phys. Rev. D}\ }\textbf {\bibinfo {volume} {42}},\ \bibinfo {pages} {2045} (\bibinfo {year} {1990})}\BibitemShut {NoStop}%
\bibitem [{\citenamefont {Domcke}\ and\ \citenamefont {Kopp}(2024)}]{Domcke2024}%
  \BibitemOpen
  \bibfield  {author} {\bibinfo {author} {\bibfnamefont {V.}~\bibnamefont {Domcke}}\ and\ \bibinfo {author} {\bibfnamefont {J.}~\bibnamefont {Kopp}},\ }in\ \href@noop {} {\emph {\bibinfo {booktitle} {58th Rencontres de Moriond on Cosmology}}}\ (\bibinfo {year} {2024})\ \bibinfo {note} {contribution to the 2024 Cosmology session of the 58th Rencontres de Moriond},\ \Eprint {https://arxiv.org/abs/2406.03244} {arXiv:2406.03244 [gr-qc]} \BibitemShut {NoStop}%
\bibitem [{\citenamefont {Kanno}\ \emph {et~al.}(2025)\citenamefont {Kanno}, \citenamefont {Soda},\ and\ \citenamefont {Taniguchi}}]{Kanno2025}%
  \BibitemOpen
  \bibfield  {author} {\bibinfo {author} {\bibfnamefont {S.}~\bibnamefont {Kanno}}, \bibinfo {author} {\bibfnamefont {J.}~\bibnamefont {Soda}},\ and\ \bibinfo {author} {\bibfnamefont {A.}~\bibnamefont {Taniguchi}},\ }\href {https://doi.org/10.1140/epjc/s10052-024-13736-z} {\bibfield  {journal} {\bibinfo  {journal} {Eur. Phys. J. C}\ }\textbf {\bibinfo {volume} {85}},\ \bibinfo {pages} {31} (\bibinfo {year} {2025})}\BibitemShut {NoStop}%
\bibitem [{\citenamefont {Wakamatsu}(2025)}]{Wakamatsu2025}%
  \BibitemOpen
  \bibfield  {author} {\bibinfo {author} {\bibfnamefont {M.}~\bibnamefont {Wakamatsu}},\ }\href {https://doi.org/10.1016/j.aop.2025.169984} {\bibfield  {journal} {\bibinfo  {journal} {Ann. Phys. (N.Y.)}\ }\textbf {\bibinfo {volume} {477}},\ \bibinfo {pages} {169984} (\bibinfo {year} {2025})},\ \Eprint {https://arxiv.org/abs/2406.18046} {arXiv:2406.18046 [quant-ph]} \BibitemShut {NoStop}%
\bibitem [{\citenamefont {Rai~Choudhury}\ and\ \citenamefont {Mahajan}(2019)}]{Rai2019}%
  \BibitemOpen
  \bibfield  {author} {\bibinfo {author} {\bibfnamefont {S.}~\bibnamefont {Rai~Choudhury}}\ and\ \bibinfo {author} {\bibfnamefont {S.}~\bibnamefont {Mahajan}},\ }\href@noop {} {\bibfield  {journal} {\bibinfo  {journal} {Phys. Lett. A}\ }\textbf {\bibinfo {volume} {383}},\ \bibinfo {pages} {2467} (\bibinfo {year} {2019})},\ \Eprint {https://arxiv.org/abs/1903.04138} {arXiv:1903.04138 [quant-ph]} \BibitemShut {NoStop}%
\bibitem [{\citenamefont {Ruggiero}(2022)}]{Ruggiero2022}%
  \BibitemOpen
  \bibfield  {author} {\bibinfo {author} {\bibfnamefont {M.~L.}\ \bibnamefont {Ruggiero}},\ }\href {https://doi.org/10.1007/s10714-022-02983-8} {\bibfield  {journal} {\bibinfo  {journal} {Gen. Relativ. Gravit.}\ }\textbf {\bibinfo {volume} {54}},\ \bibinfo {pages} {97} (\bibinfo {year} {2022})}\BibitemShut {NoStop}%
\end{thebibliography}
%

\end{document}